\documentclass[twocolumn,showpacs,pre, floatfix]{revtex4}
\ifx\pdfoutput\undefined
\usepackage{graphicx}
\else
\usepackage[pdftex]{graphicx}
\usepackage{epstopdf}
\fi

\usepackage[center]{subfigure}

\begin{document}

\title{Steady states and linear stability analysis of precipitation pattern formation at geothermal hot springs}

\author{Pak Yuen Chan and Nigel Goldenfeld}
\affiliation{Department of Physics, University of Illinois at
Urbana-Champaign, Loomis Laboratory of Physics, 1110 West Green
Street, Urbana, Illinois, 61801-3080.}

\begin{abstract}

A dynamical theory of geophysical precipitation pattern formation is
presented and applied to irreversible calcium carbonate (travertine)
deposition. Specific systems studied here are the terraces and domes
observed at geothermal hot springs, such as those at Yellowstone
National Park, and speleothems, particularly stalactites and
stalagmites.  The theory couples the precipitation front dynamics with
shallow water flow, including corrections for turbulent drag and
curvature effects.  In the absence of capillarity and with a laminar
flow profile, the theory predicts a one-parameter family of steady
state solutions to the moving boundary problem describing the
precipitation front.  These shapes match well the measured shapes near
the vent at the top of observed travertine domes. Closer to the base of
the dome, the solutions deviate from observations, and circular
symmetry is broken by a fluting pattern, which we show is associated
with capillary forces causing thin film break-up.  We relate our model
to that recently proposed for stalactite growth, and calculate the
linear stability spectrum of both travertine domes and stalactites.
Lastly, we apply the theory to the problem of precipitation pattern
formation arising from turbulent flow down an inclined plane, and
identify a linear instability that underlies scale-invariant travertine
terrace formation at geothermal hot springs.

\end{abstract}

\pacs{05.45.Ra, 87.23.n, 47.54.-r, 89.75.Kd, 47.20.Hw, 47.15.gm,
47.55.np} \maketitle

\section{Introduction}
Geophysical pattern formation concerns how geological patterns and
landscapes are formed as a result of the underlying physical and
chemical dynamics. The aim is to predict the static, dynamical and
statistical properties of the variety of geological structures formed.
Recently studied examples include travertine motifs, namely
dams\cite{WOOD91}, domes\cite{GOLD06} and
terraces\cite{FOUK00,FOUK01,HAMM06,GOLD06},
stalactites\cite{SHOR05a,SHOR05b}, as well as that of other patterns
such as sand dunes\cite{PYE90,LANC96}, black smoker chimneys at
hydrothermal vents\cite{kerr1996cag}, columnar
joints\cite{goehring2006eis} and braided river
networks\cite{murray1994cmb}.

This paper focuses on the formation of travertine structures near
geothermal hot springs.  In such systems, hot spring water emerges from
a vent, and deposits calcium carbonate as a mineral generally termed
travertine as it degasses carbon
dioxide\cite{FOUK00,FOUK01,GOLD06,WOOD91}. The formation of stalactites
in limestone caves, which are also caused by carbonate precipitation,
will also be briefly discussed.

\begin{figure}
\includegraphics[width=0.99\columnwidth]{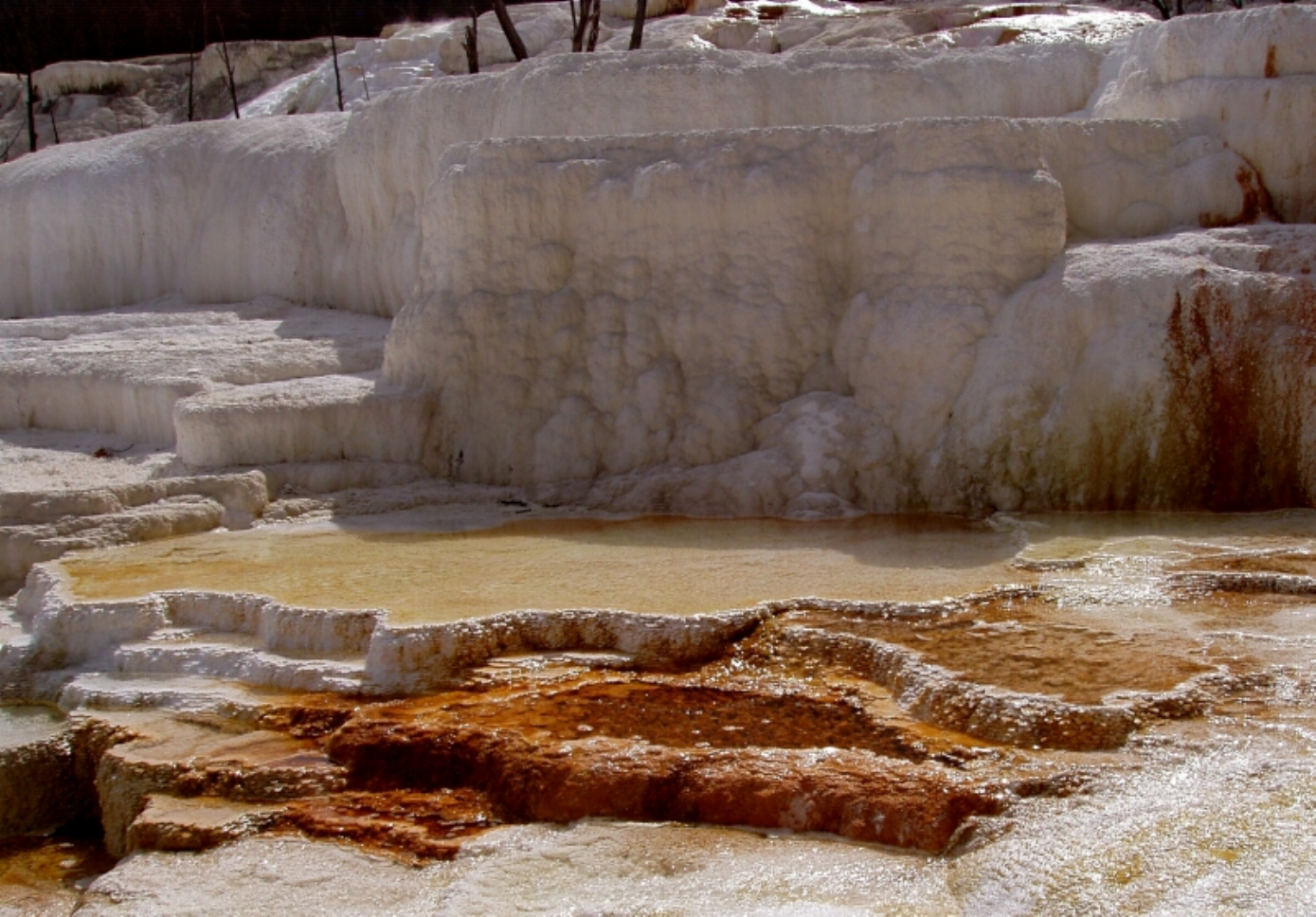}
\caption{(Color online) Travertine formation at Angel Terrace,
Mammoth Hot Springs, WY, showing a large pond, of order 1 meter in
diameter, and smaller features.}\label{AT3}
\end{figure}

The majority of the work done on the subject has focused on the
microscopic aspects of the problem, such as the role of
biomineralization due to thermophilic microbes\cite{FOUK00, FOUK01},
the CO$_2$ degassing mechanisms\cite{HERM87, ZHAN01}, mineral
compositions\cite{BARN71, CHAF91} and crystal structure\cite{BUSE86,
RENA96}.  Here we are interested in the formation of
macroscopic structures and motifs, such as domes, stalactites, and
terraces\cite{GOLD06}, which are universal, {{\it i.e.}}, independent
of microscopic details.  In addition, we are interested in the
resulting patterns and their correlations, rather than absolute rates
of growth; accordingly, microscopic mechanisms that contribute to
kinetics, including nucleation processes and potential
biomineralization effects, are present in our work through the
choice of time scale.  There are no extra terms in the equations of
motion whose presence can be attributed specially to any one of these
microscopic processes.

There are two principal mathematical difficulties encountered in
studying these macroscopic structures. First, the problem is highly
nonlinear.  As the carbonate is precipitated onto the surface, the
surface evolves, which then changes the flow path of the fluid, thus
affecting how precipitation takes place. This interplay between fluid
flow and surface growth leads to a moving-boundary problem, which is
mathematically difficult to solve. Second, the problem involves a
variety of depositional processes, including solute advection, a
complex sequence of chemical reactions, CO$_2$ degassing, as well as
mass transfer between a solid and a liquid. Given that each of these
processes is complicated and non-trivial to model on its own, a
holistic approach capturing all of them would not be mathematically
tractable.

The purpose of this paper is to explore a simplified mathematical
formulation of this problem that captures the essential large-scale
dynamics.  Because of the complexity of the problem, the resulting
equations are very complicated, making it difficult, if not impossible,
to understand the whole flow system using this approach.  It turns out,
however, that the equations can be solved analytically under some
simple situations, where symmetry can be exploited and simplifications
can be made. The formations of domes\cite{GOLD06} and
stalactites\cite{SHOR05a,SHOR05b} are examples of such situations, as
is the pioneering work of Wooding on travertine dams\cite{WOOD91}.  In
these systems, there is a thin film of fluid flowing over the motif in
a laminar fashion (in the case of domes and stalactites, for example).
We will see that these simple motifs are straightforward to calculate
in the case that capillary forces can be neglected.  If the fluid film
becomes too thin, due to its spreading over the surface, contact lines
can be formed, resulting in rivulets and the breaking of pure
rotational symmetry.  In the case of domes, this is manifested in a
fluting pattern near the base of the dome\cite{GOLD06}.  Such effects
are difficult to include analytically, although we have previously
shown that they can be captured correctly using a cell dynamical system
model\cite{GOLD06}, and this is discussed in more detail below.

Although we cannot use this analytical theory to study the detailed
shapes of the complex landscape of ponds and terraces, we are able to
expose the dynamical linear instabilities, whose evolution into the
nonlinear regime give rise to the landscape.  We will see that the
linear stability spectrum, in the absence of capillarity effects,
always predicts a positive growth rate.  The absence of a length scale
arising in this calculation suggests that the actual landscapes might
be scale invariant, a conclusion that is reinforced by our studies of
the statistical properties of these landscapes using our cell dynamical
system model and photographic evidence\cite{VEYS07, GOLD06}

The study reported here is a complement to our simulation
work\cite{GOLD06, VEYS06, VEYS07} implemented as a cell dynamical
system. This model has been shown to be capable of describing the
actual dynamics\cite{GOLD06}, not only in the simple cases where the
analytical approach is successful, but also in the fully nonlinear
regime. For example, it has been shown that this cellular model
generically gives rise to a complex, terraced landscape, which is
similar to the one observed in the field.  The cellular model also
makes detailed predictions for the landscape statistics, including the
pond area distribution and the distribution of pond anisotropy.  In
addition, the model successfully predicts that the main mode of pond or
terrace growth is uphill pond inundation, a result confirmed by
time-lapse photographic studies.

Although seemingly different, both the analytical approach and the cell
dynamical system approach incorporate the same physics, and so should
be expected to yield identical predictions.  In \cite{GOLD06} this was
tested, by using the cellular model to solve the problem of dome
formation.   The analytical theory in the absence of surface tension
cannot account for the fluting seen away from the vent of domes,
because the fluting arises from contact line formation.  The analytical
theory for domes, as we will discuss in detail below, contains one
parameter that sets the scale for the patterns: this scale factor $r_0$
is a combination of the upward growth velocity, the mass transfer
coefficient describing how material is incorporated into the growing
substrate, the flux of water emerging from the vent, the gravitational
acceleration and the fluid viscosity.  When surface tension effects are
included, the capillary length $d_0$ must also be included.  Thus, our
theory is a two parameter theory for the entire range of travertine
depositional phenomena.  The analytical theory can be used to predict
the position on the dome at which capillary effects become
important: this must occur at a location that is independent of the
ratio $r_0/d_0$, and hence this critical angle has a prescribed
dependence on the underlying parameters which enter into the formula
for $r_0$.  This prediction, arising from the analytical theory, was
verified to occur also in the computer simulations of the cellular
model\cite{GOLD06}.  As a result, we conclude that the two formulations
are indeed equivalent, and may be used interchangeably depending on
which is more suited to the problem at hand.

This paper is organized as follows.  In Section \ref{model}, we derive
the equations governing the dynamics of fluid flow coupled to the
moving boundary problem describing travertine precipitation.  Section
\ref{domes} describes the circularly symmetric solutions of these
equations, and presents the linear stability analysis of the steady
state uniformly translating solutions.  We compare our analysis to a
similar one\cite{SHOR05a,SHOR05b} that describes the shapes of
stalactites in Section \ref{stalactite} and compute the linear
stability spectrum of these structures too.  We turn in Section
\ref{sec_damming} to a study of turbulent flow down an inclined plane,
and calculate the linear stability spectrum for the coupled flow and
moving boundary problem, exposing the linear instability that is at the
heart of the terraced landscape architecture.  We conclude in Section
\ref{sec_conclusion}.

\section{Model for Precipitation Pattern Formation}
\label{model}

We consider a stream of water flowing over a terrain, from which
calcium carbonate is then, due to geochemical processes to be discussed
below, precipitated onto the landscape. The landscape is thus
constantly changing in response to the fluid flow. This change of
landscape, in turn, affects the flow path of the fluid, which than
influences how subsequent precipitation takes place. We derive the
governing equations describing both fluid flow and surface growth.  We
first focus on the surface growth, and related precipitation dynamics,
and then move onto the fluid flow.  These two aspects will be combined
to provide the complete description of the system.

\subsection{Surface Growth}
\begin{figure}[t]
\includegraphics[width=0.99\columnwidth]{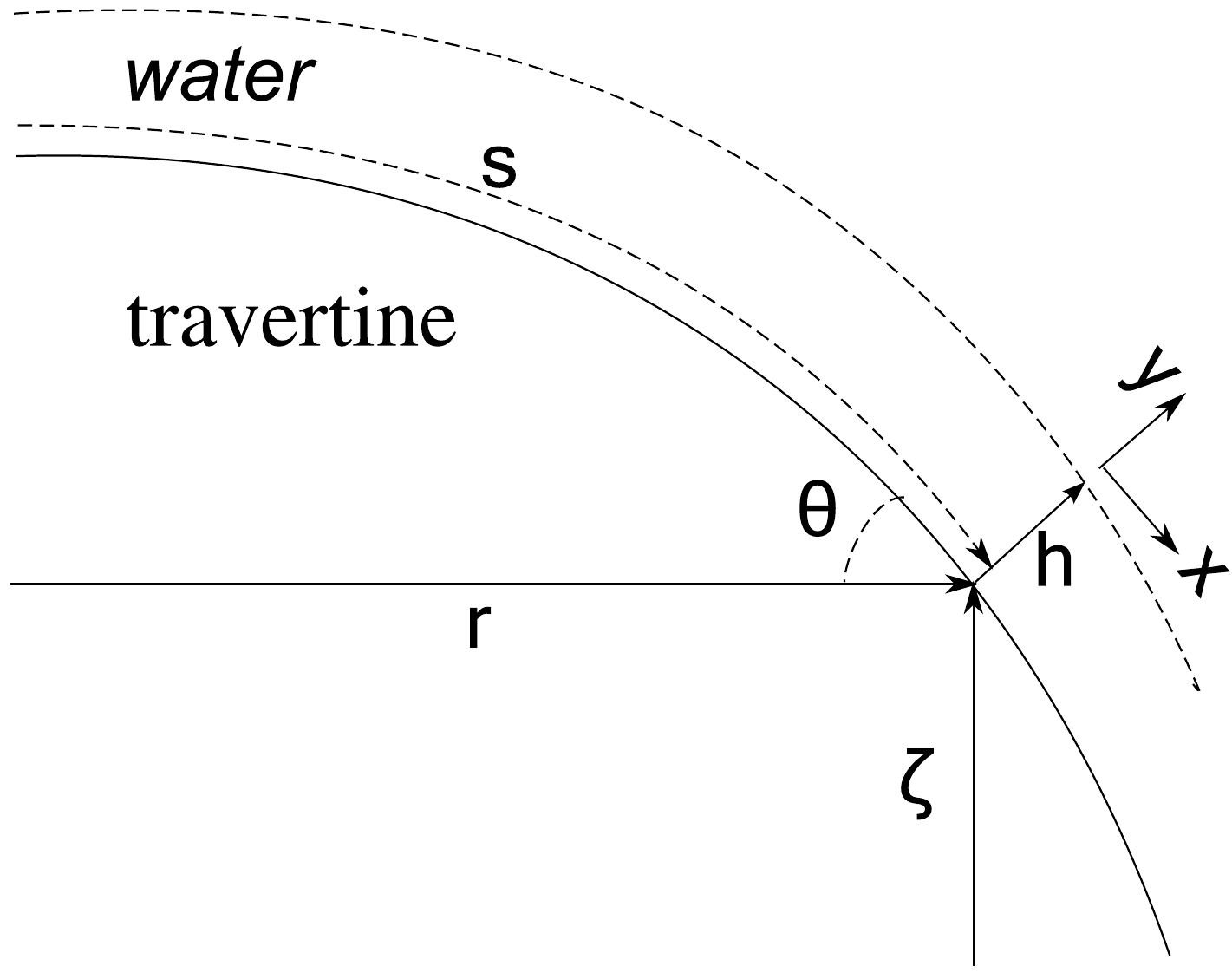}
\caption{The coordinate system for the model of fluid flow coupled to
precipitation moving boundary dynamics.} \label{fig_coordinate}
\end{figure}

A surface can generally be characterized by the local curvature,
$\kappa$.  In one dimension, or in cases where symmetry reduces the
system to be effectively one dimensional, $\kappa$ is defined by
\begin{equation}
\kappa=\frac{\partial\theta}{\partial s},
\end{equation}
where $\theta$ is an angle between the local tangent of the curve and a
fixed axis, and $s$ is the arc length measured from some fixed point on
the curve, as shown in Fig. (\ref{fig_coordinate}).  If the normal
velocity $v_n$ of the surface is prescribed everywhere, then the
evolution of the curvature follows the kinematic
equation\cite{brower1983gam, BENJ83, BROW84}:
\begin{equation}
\frac{\partial\kappa}{\partial t}\bigg|_{\theta} = -\kappa^2\left(1+
\frac{\partial^2}{\partial \theta^2}\right)v_n, \label{growth_eqn}
\end{equation}
The time derivative in the equation is defined with respect to fixed
$\theta$.  The first term in Eq. (\ref{growth_eqn}) describes the
change in $\kappa$ due to the change in the overall scale of the
object, whereas the second term describes the change in $\kappa$ at a
point due to the difference in growth rates in the neighborhood of that
point.

Eq. (\ref{growth_eqn}) is purely geometrical; for any given function
$v_n$, the evolution of $\kappa$ is determined.  So, physics enters in
constructing a realistic and mathematically tractable model for $v_n$,
which, in the case considered here, depends on water chemistry, surface
kinematics, chemical advection and fluid flow state.  In carbonate
systems, in additional to the CaCO$_3$ concentration, precipitation is
mainly controlled by the CO$_2$ concentration (partially reflected in
the measured pH), which is also influenced by its temperature-dependent
solubility in the fluid. As the pH increases or the temperature
decreases, the solvability of CaCO$_3$ decreases and supersaturated
CaCO$_3$ will be precipitated onto the surface. While the decrease in
temperature is mainly due to heat loss to the environment, the increase
in pH is due to the loss of CO$_2$ by a variety of outgassing
mechanisms\cite{HERM87, ZHAN01}.  Although the detailed water chemistry
and depositional processes are quite complicated, for the purposes of
the present work, it suffices to use a simplification of the governing
chemical reactions: Ca$^{2+}$ + 2HCO$_3^-$ $\rightleftharpoons$
CaCO$_3$(s) + H$_2$O + CO$_2$(g).  In summary, the system tends to
produce more CaCO$_3$ as CO$_2$ concentration decreases through
outgassing.

Mass transfer between a fluid and a solid is a complicated
problem\cite{CAMP82, CAMP83, WOOD91}; these nontrivial chemical
processes only make it harder.  A complete description of the
precipitation dynamics, which will give us the normal growth velocity
$v_n$, involves writing down, in addition to the fluid dynamics
equations, advection-reaction-diffusion equations for each chemical and
appropriate boundary conditions. Short {\it et
al.}\cite{SHOR05a,SHOR05b} followed this approach in the study of
stalactite formation.  What they found, after solving all these
equations and taking limits appropriate for the timescales of interest
to them, is that $v_n$ is proportional to the local fluid thickness,
$h$, with all the chemistry entering only into the proportionality
constant.

A simple interpretation of this result can be obtained by studying
the scales of processes involved in stalactite formation, using parameter
values from Ref. \cite{SHOR05b}.  The fluid flow
is a laminar flow, with Reynold's number about $0.01-1$ .  The
thickness of the flow, $h$, is typically on the order of 10$\mu$m.
The time scale for the concentration of CaCO$_3$ to equilibrate
across the layer is thus $h^2/D\sim 0.1$s, where $D$ is the
diffusion constant. Next, the traversal time, the time for a parcel
of fluid to flow along the stalactites, is about $100$s. Because
only $1$ percent of the total CaCO$_3$ mass is precipitated
throughout the flow, we can assume that the CaCO$_3$ concentration,
and thus the pH, are uniform both across the fluid layer and along
the stalactite. The temperature can also be assumed to be constant
since the fluid is so thin.  The precipitation rate is then
controlled only by the CaCO$_3$ available, which is proportional to
the thickness of the fluid.

In other carbonate systems, such as at travertine-forming hot springs,
this relation between $v_n$ and $h$ does not hold simply due to the
fact that the fluid thickness is larger, and the velocity is larger; as
a result a turbulent boundary layer is formed near the precipitation
front.  What happens outside the boundary layer is too distant to
affect precipitation near the boundary.  In a turbulent flow, instead
of depending on $h$, the precipitation front velocity $v_n$ depends on
the fluid velocity\cite{CAMP82,CAMP83}. Wooding\cite{WOOD91}, in the
study of steady-state dam formation, took this into consideration and
arrived at the conclusion that $v_n$ is directly proportional to the
depth-averaged tangential fluid velocity, $U$, {\it i.e.}
\begin{equation}
v_n=GU,
\label{vn_eqn}
\end{equation}
where $G$ is a mass transfer coefficient depending on water chemistry
and spectral features of the turbulent flow\cite{CAMP82,CAMP83}.  For
present purposes, the functional form of $G$ is not of interest: we
shall treat it as a phenomenological parameter, and as we shall see,
its role in the theory developed here is to contribute to the
characteristic length scale $r_0$ of patterns.

To summarize: all the details of water chemistry, including
supersaturation, outgassing, solute diffusion, fluid turbulence,
temperature and pH, which on their own are complicated processes and
are nontrivial to model, enter into the picture only through a mass
transfer coefficient, $G$.  In principle, $G$ may exhibit spatial
fluctuations; however, we shall assume that these are on a scale small
compared to the features we are describing, and thus we will consider
$G$ to be a constant locally along the flow path.  Over the entire
geothermal spring system, it is possible that there will be a small
spatial variation in the mean value of $G$, but the weak dependence of
$G$ on governing parameters\cite{CAMP82, CAMP83, WOOD91} strongly
suggests that this can reasonably be neglected.

\subsection{Fluid Dynamics}
A complete description of incompressible fluid dynamics is given by
the Navier-Stokes equation
\begin{equation}
\frac{\partial \vec{u}}{\partial t} +
\vec{u}\cdot\nabla\vec{u}=\frac{-1}{\rho}\nabla P  +
\nu\nabla^2\vec{u} + \vec{g}, \label{NS_eqn}
\end{equation}
with $\vec{\nabla}\cdot\vec{u}=0$ for incompressibility, no-slip and
stress-free boundary conditions at the solid-liquid and liquid-gas
interfaces, respectively, where $\vec{u}$, $\rho$, $P$, $\nu$ and
$g$ are the fluid velocity, density, pressure, viscosity and
gravitational acceleration.  We will use the Poiseuille solutions of
the Navier-Stokes equations for domes, where the flow is laminar, but
for turbulent flows, such as those which form the travertine terraces,
we will employ a depth-averaging approximation, in conjunction with the
Ch\'ezy approximation\cite{CHEZ76} for hydraulic friction.

Since the spatial scale over which the landscape changes is usually
much larger than the fluid thickness, {\it i.e.} $h\kappa\ll 1$, we
can make use of the shallow water approximation and expand Eq.
(\ref{NS_eqn}) in powers of $h\kappa$.  If we take $\kappa$ to be
zero, we arrive at the de Saint-Venant equation\cite{VENA71}
\begin{equation}
\frac{\partial (Uh)}{\partial t}+\frac{\partial (U^2h)}{\partial s} =-gh\frac{\partial h}{\partial s} + gh\sin\theta - \frac{C_fU^2}{h}
\label{dSV_eqn}
\end{equation}
with equation of continuity
\begin{equation}
\frac{\partial h}{\partial t} + \frac{\partial (Uh)}{\partial s} = 0
\end{equation}
where $C_f$ is the Ch\'ezy coefficient\cite{CHEZ76}, which empirically
describes the energy lost due to turbulence, in a manner consistent
with Kolmogorov's 1941 scaling theory of turbulence
(K41)\cite{KOLM41,SREE99}, and $s$ is the arc length measure from a
reference point at the top, as shown in Fig. (\ref{fig_coordinate}).

The de Saint-Venant equation only holds on flat surfaces.  When the
surface grows, flow instabilities trigger various patterns to form;
and the de Saint-Venant equation is no longer valid.  For a general
curved surface, the Dressler equation\cite{DRES78,SIVA81} has to be
used:
\begin{equation}
\frac{1}{g}\frac{\partial u_0}{\partial t} + \frac{\partial E}{\partial s} = \frac{-C_fu^2}{gh(1-\kappa h/2)}
\label{Dressler_eqn_1}
\end{equation}
\begin{equation}
(1-\kappa h) \frac{\partial h}{\partial t} +\frac{\partial q}{\partial s} = 0
\end{equation}
where
\begin{equation}
u(s,n,t)=\frac{u_0(s,t)}{1-\kappa n},
\end{equation}
\begin{equation}
E(s,t)=\zeta+h\cos\theta+\frac{p_h}{\rho g}+\frac{u_0^2}{2g (1-\kappa h)^2},
\end{equation}
\begin{equation}
q(s,t)=-\frac{u_0}{\kappa}\log(1-\kappa h).
\label{Dressler_eqn_2}
\end{equation}
where $\zeta$ is the height of the underlying surface measured from
a fixed horizontal axis, as shown in Fig. (\ref{fig_coordinate}),
$p_h$ is the pressure head at the fluid surface, $\rho$ is the fluid density,
$E$ is the energy density and $q$ is the local flux. When $\kappa$ is
set to zero and $\theta$ is small, the Dressler equations reduce to those of de Saint-Venant.

As we have seen, the way fluid flows depends on the landscape it is
flowing over, which itself is evolving over time.  Now, Eq.
(\ref{Dressler_eqn_1})-(\ref{Dressler_eqn_2}) (or Eq. (\ref{NS_eqn})) and Eq. (\ref{vn_eqn})
describe these two dynamics, respectively.  However, we do not have to
consider both dynamics on the same footing because there is a
separation of time scales; the rate of fluid flow is on the order of
cm/sec, but the rate of precipitation is on much slower geological
scales.  The latter is on the order of 1 mm/day and 1 cm/century in the
cases of Yellowstone travertines\cite{FRIE70,PENT90, FOUK00} and
stalactites\cite{SHOR05b}, respectively. Accordingly the fluid flow
responds quickly to the landscape, but the landscape responds extremely
slowly to the fluid flow.  We can then assume that the fluid flow is in
its steady state when we discuss the landscape evolution; {\it i.e.},
we can drop all the time derivatives in the fluid flow equations.  This
quasi-stationary model will now be used to study the steady states of
a variety of geological motifs and their stabilities.

\section{Travertine domes}
\label{domes}

\subsection{Steady state}
\label{Travertine_domes_steady_state}
\begin{figure}
\includegraphics[width=0.99\columnwidth]{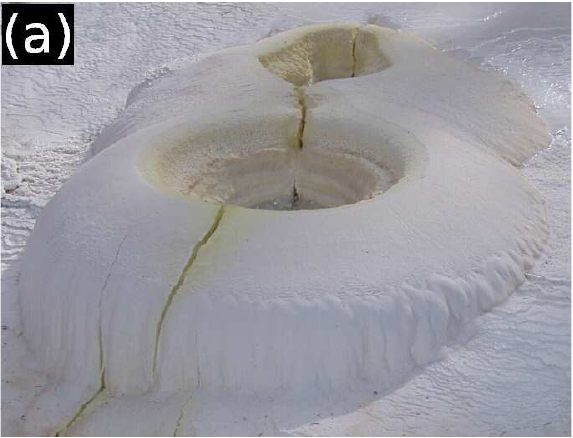}
\includegraphics[width=0.99\columnwidth]{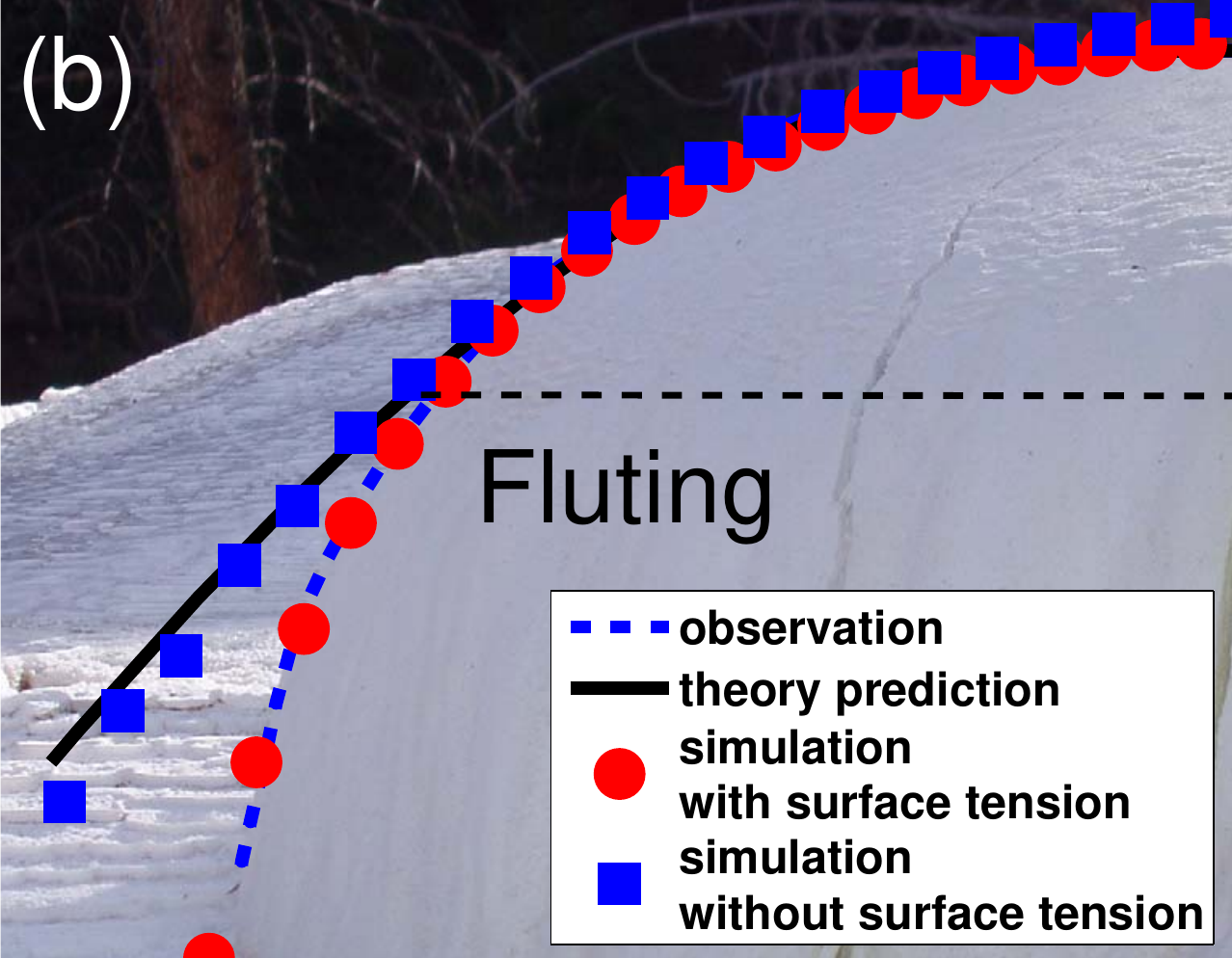}
\caption{(Color online) Travertine dome at Mammoth Hot Springs, WY. (a)
Dome whose central pond is 50.3cm in diameter. (b) Dome profile
compared with theory and simulation of Ref. \cite{GOLD06}.  The black
curve is the analytical prediction from Eq. (\ref{dome_shape}), using the
value $r_0=43 cm$.  The
red filled circles show the profile of a simulated dome, including the
effects of surface tension.  The blue dashed line is a consensus dome
profile generated by averaging the dome shown with one other field
observation.  The blue filled squares show the profile of a simulated
dome without surface tension\cite{GOLD06}.}
\label{Dome}
\end{figure}

Our first example is the circularly symmetrical domes found in
Yellowstone National Park, as shown in Fig. (\ref{Dome}a).  A number of
approximations and simplifications can be made before we proceed.
First, the growth rate of these domes is on the order of $1-5$mm/day
and the fluid flow rate is on the order of $1$mm/s, so we have a
separation of time scales. Second, our field observations indicate
that the thickness of the fluid film flowing over the domes is very
small compared to the curvature of the surface;  thus, we make the
approximation that the fluid is flowing down a (locally) constant
slope.  Third, as suggested by the field observations, the domes
have a high degree of circular symmetry, so we can assume the
solution to be circularly symmetrical and focus only on the radial
part of the solution, which is effectively one dimensional.  Fourth,
the flow is apparently laminar, so we can use the Poiseuille-Hagen
profile for the velocity in thin film:
\begin{equation}
u(y)=\frac{gh^2\sin\theta}{2\nu}
\left[2\frac{y}{h}-\left(\frac{y}{h}\right) ^2\right],
\label{P-H_profile}
\end{equation}
where $\theta$ is the slope of the surface and $y$ is the transverse
coordinate, as shown in Fig. (\ref{fig_coordinate}).  By assuming
circular symmetry, $h$ can be related to the axial distance from the
vent, $r$, by the conservation of fluid volume:
\begin{equation}
Q=2\pi r\int_0^h u(y)dy=\frac{2\pi gr h^3 \sin\theta}{3\nu},
\label{mass_conservation}
\end{equation}
where $Q$ is the total volumetric flux coming out of the vent.  Eq.
(\ref{P-H_profile}) and (\ref{mass_conservation}) can be combined to
give
\begin{equation}
U \equiv \frac{1}{h}\int_0^h u(y)dy=\left(
\frac{\alpha\sin\theta}{r^2} \right)^{1/3}, \label{averaged_v}
\end{equation}
where $\alpha\equiv g Q^2/12\pi\nu$.  We will see later that the
assumption of laminar flow is self-consistently verified.  Putting
Eq. (\ref{averaged_v}) into Eq. (\ref{growth_eqn}) and using Eq.
(\ref{vn_eqn}), gives
\begin{equation}
\frac{\partial\kappa}{\partial t}\bigg|_{\theta} = -\kappa^2\left[1+
\frac{\partial^2}{\partial \theta^2}\right]G\left(
\frac{\alpha\sin\theta}{r^2} \right)^{1/3}. \label{dome_eqn}
\end{equation}
This is the governing equation for the dome profile.  Suggested by
the shape of the dome, we seek a solution which steadily
translates upwards without a change of shape, {\it i.e.},
$\partial_t\kappa|_{\theta}=0$, with velocity $v_t$. Eq.
(\ref{dome_eqn}) gives
\begin{equation}
G\left(\frac{\alpha\sin\theta}{r^2}\right)^{1/3} = v_t\cos\theta,
\end{equation}
Rearranging terms gives the shape of the dome as a one-parameter
family of curves
\begin{equation}
\frac{r(\theta)}{r_0} = \sqrt{\frac{\sin\theta}{\cos^3\theta}},
\label{dome_shape}
\end{equation}
where the scale factor $r_0\equiv \sqrt{G^3\alpha/v_t^3}$.  Eq.
(\ref{dome_shape}) is plotted in Fig. (\ref{Dome}b).  Good agreement
is obtained between our theory and the observations below a critical
angle $\theta_c$. From the fit, and the typical parameter values
$G\sim 10^{-8}$, $v_t\sim 1\mathrm{mm/day}$ and $Q\sim
1\mathrm{cm}^3\mathrm{/sec}$, we obtain $U\sim 25 \mathrm{mm/sec}$
and $h\sim1-10\mathrm{mm}$, and a Reynolds number,
$\mathrm{Re}\equiv Uh/\nu\sim 10-100$. The assumption of laminar
flow is self-consistently verified.

The agreement between this analysis and observation shows that the
growth of the dome is mainly determined by the geometry, because the
only $r$ dependence enters through the mass conservation, which is
determined by geometry.  To see this, suppose that the dome was a one
dimensional object.  Then, the mass conservation equation, Eq.
(\ref{mass_conservation}), would become $Uh=q_0$, for some constant
flux $q_0$, without any $r$ dependence.  Under the same approximation
of local flatness, the final equation for $U$, Eq. (\ref{averaged_v}),
would thus be independent of $r$.  We would then not be able to solve
for $r$ by substituting $U$ into Eq. (\ref{growth_eqn}).  In this case,
we would have to solve the equations without using the locally-flat
approximation. In other words, the fact that we can ignore the details
of the flow, by assuming local flatness, to obtain the shape of the
domes implies that geometry plays a more important role than fluid flow
in the formation of domes.

For angles $\theta > \theta_c$, the analytical profile deviates from
our field photograph.  The point of deviation is associated with an
apparent change in the dome morphology, with a fluting pattern
superimposed on the dome profile.  This is due to the effects of
surface tension at the air-water-travertine interface\cite{GOLD06}.
Instead of covering the whole surface uniformly, the fluid separates
and covers only a fraction of the surface.  Along the wetted surface,
the regular growth law still applies and thus the surface grows, until
a point at which the difference in heights between the wetted and dry
surfaces is so large that the flow changes its path to flow along the
dry surface. This process repeats itself and, on average, results in a
slower growth when compared with a uniformly-wetted dome, so the
theoretical prediction should be higher than the observation for
$\theta > \theta_c$, as seen in Fig. (\ref{Dome}b). The analytical
solution for the dome profile neglects surface tension, but leads to a
prediction for the scaling dependence of the critical angle on the
model parameters\cite{GOLD06}.

It is not trivial to include surface tension in our analytical model,
but its effect can be examined by using the cellular model, in which
one can switch on and off surface tension. Fig. (\ref{Dome}b),
reproduced from Ref. \cite{GOLD06} shows the prediction of dome shapes
from the cellular model with and without surface tension.  It is clear
that by appropriate choice of $d_0$ the simulation result coincides
with the observation when surface tension is present, and agrees with
the analytical prediction otherwise. This is direct evidence for the
effect of surface tension near fluting.

For completeness, we mention that this is not an artifact of having
\lq\lq enough fitting parameters to fit an elephant".  In Ref.
\cite{GOLD06} was presented a scaling argument for the critical angle
at which capillary effects become important. The inclusion of surface
tension introduces an additional length scale, namely, the capillary
length, $d_c$, into the problem. Now, the only other length scale in
the problem is $r_0= \sqrt{gG^3Q^2/\nu v_t^3}$.  Since $\theta_c$ is
dimensionless, it can only depend on the ratio $r_0/d_c$ and $G$.  For
a given chemical environment, $G$ is fixed and we are left with the
prediction, derived from our analytical solution, that
\begin{equation}
\theta_c= \hat{f}\left(\frac{\sqrt{gQ^2/\nu v_t^3}}{d_c}\right),
\label{contact_line}
\end{equation}
where $\hat{f}(x)$ is a scaling function.  This data collapse, which predicts $\theta$ depends not on the parameters separately, but only on the combination $\sqrt{(gQ^2/\nu
v_t^3)}/d_c$, was verified using our discrete cellular model\cite{GOLD06}, wherein the form of $\hat{f}(x)$ was calculated.

\subsection{Linear stability analysis}
\label{travertine_domes_linear_stability}

To complete the analysis, we study the stability of the solution,
Eq. (\ref{dome_shape}).  By following the approach Liu and Goldenfeld used in studying the linear stability of dendritic solidification\cite{Liu88}, we consider a perturbed solution, $r(\theta) = \bar{r}(\theta) + \delta r(\theta) e^{\lambda t}$, where
$\bar{r}(\theta)$ is the solution in Eq. (\ref{dome_shape}) and $\delta r$ is a perturbation.  Substituting this into the governing equation, Eq. (\ref{dome_eqn}), and expanding in $\delta r$, we obtain
\begin{equation}
\lambda \frac{d\delta r}{d\theta} +
\frac{2G\alpha^{1/3}\cos\theta}{3}\left[1+\frac{d^2}{d\theta
^2}\right] \frac{\delta r\sin^{1/3}\theta}{\bar{r}^{5/3}}=0,
\label{dome_ev_eqn}
\end{equation}
where the boundary conditions are
\begin{equation}
\delta r(0)=0, \qquad\delta r\left(\frac{\pi}{2}\right) = 0,
\label{eqn_bc_sym}
\end{equation}
for symmetric modes and,
\begin{equation}
\frac{d(\delta r(0))}{d\theta}=0, \qquad\delta r\left(\frac{\pi}{2}\right) = 0,
\end{equation}
for antisymmetric modes.  This is an eigenvalue problem
and the spectrum tells us the stability of the solution.  It is
sufficient to examine the asymptotic behaviors of
$\delta r$ for different values of $\lambda$ to extract sufficient
information about the stability.  Expanding Eq. (\ref{dome_ev_eqn}) in
small $\theta$ gives
\begin{equation}
\frac{d^2\delta r}{d\theta^2}-\frac{1}{\theta}\frac{d\delta r}{d\theta}
+\frac{3}{4\theta ^2}\delta r = 0,
\end{equation}
which is independent of $\lambda$ and which possesses power-law solutions of the form $\delta r
\sim \theta^{1/2}, \theta^{3/2}$.  These correspond to the symmetric and antisymmetric
modes respectively.

The asymptotic behavior in the opposite limit can be studied by making the transformations $g(\theta) = \delta r(\theta)\sqrt{\cot\theta}$ and $x=\tan\theta$,
which results in
\begin{equation}
\frac{d^2 g(x)}{dx^2}+p(x)\frac{dg(x)}{dx}+q(x)g(x)=0,
\end{equation}
where
\begin{equation}
p(x)=\lambda '\sqrt{x(1+x^2)}-\frac{2x}{1+x^2},
\end{equation}
\begin{equation}
q(x)=\frac{\lambda '\sqrt{1+x^2}}{2\sqrt{x}}+\frac{2x^2-1}{(1+x^2)^2},
\end{equation}
and,
\begin{equation}
\lambda '\equiv \frac{3\alpha^{1/6} G^{3/2}\lambda}{2v_t^{5/2}}.
\end{equation}
The asymptotic behaviors of these functions, as $x\rightarrow +\infty$, are
\begin{equation}
p(x) \sim \lambda' x^{3/2} + \frac{\lambda'}{2x^{1/2}} - \frac{2}{x} + O\left(\frac{1}{x^{5/2}}\right),
\end{equation}
and,
\begin{equation}
q(x) \sim \frac{\lambda' x^{1/2}}{2} + \frac{\lambda'}{4x^{3/2}} + \frac{2}{x^2}+O\left(\frac{1}{x^{7/2}}\right).
\end{equation}
The asymptotic behavior of $g(x)$ as $x \rightarrow +\infty$, for positive values of $\lambda
'$, can be computed by defining $g(x) \equiv \exp(S(x))$, where $S(x)$ satisfies
\begin{equation}
\frac{d^2S}{dx^2} + \left(\frac{dS}{dx}\right)^2 + p(x)\frac{dS}{dx} + q(x) = 0.
\label{eqn_dome_asym}
\end{equation}
Using the eikonal approximation that $S''(x) \ll (S'(x))^2$, which is valid for $x \rightarrow +\infty$, Eq. (\ref{eqn_dome_asym}) can be solved asymptotically to give the two linearly independent solutions
\begin{equation}
S_1(x) \sim \frac{-2\lambda'}{5}x^{5/2}-\lambda'x^{1/2} + \ln(x),
\end{equation}
and,
\begin{equation}
S_2(x) \sim \frac{-1}{2}\ln(x),
\end{equation}
which are equivalent to,
\begin{equation}
g_1(x) \sim \frac{1}{x}\exp\left(\frac{-2\lambda '}{5}x^{5/2}-\lambda'x^{1/2}\right),
\label{eqn_dome_stability_large_asym}
\end{equation}
and,
\begin{equation}
g_2(x) \sim \frac{1}{\sqrt{x}} +\frac{3}{2\lambda x^3} -\frac{7}{4\lambda x^5} +O\left(\frac{1}{x^{11/2}}\right),
\label{eqn_dome_stability_large_asym2}
\end{equation}
where a series expansion in the form of,
\begin{equation}
g_2(x)=\frac{1}{\sqrt{x}}\sum_{n=0}^\infty \frac{a_n}{x^{n/2}},
\end{equation}
is performed to arrive at Eq. (\ref{eqn_dome_stability_large_asym2}).

\begin{figure}
\includegraphics[width=0.99\columnwidth]{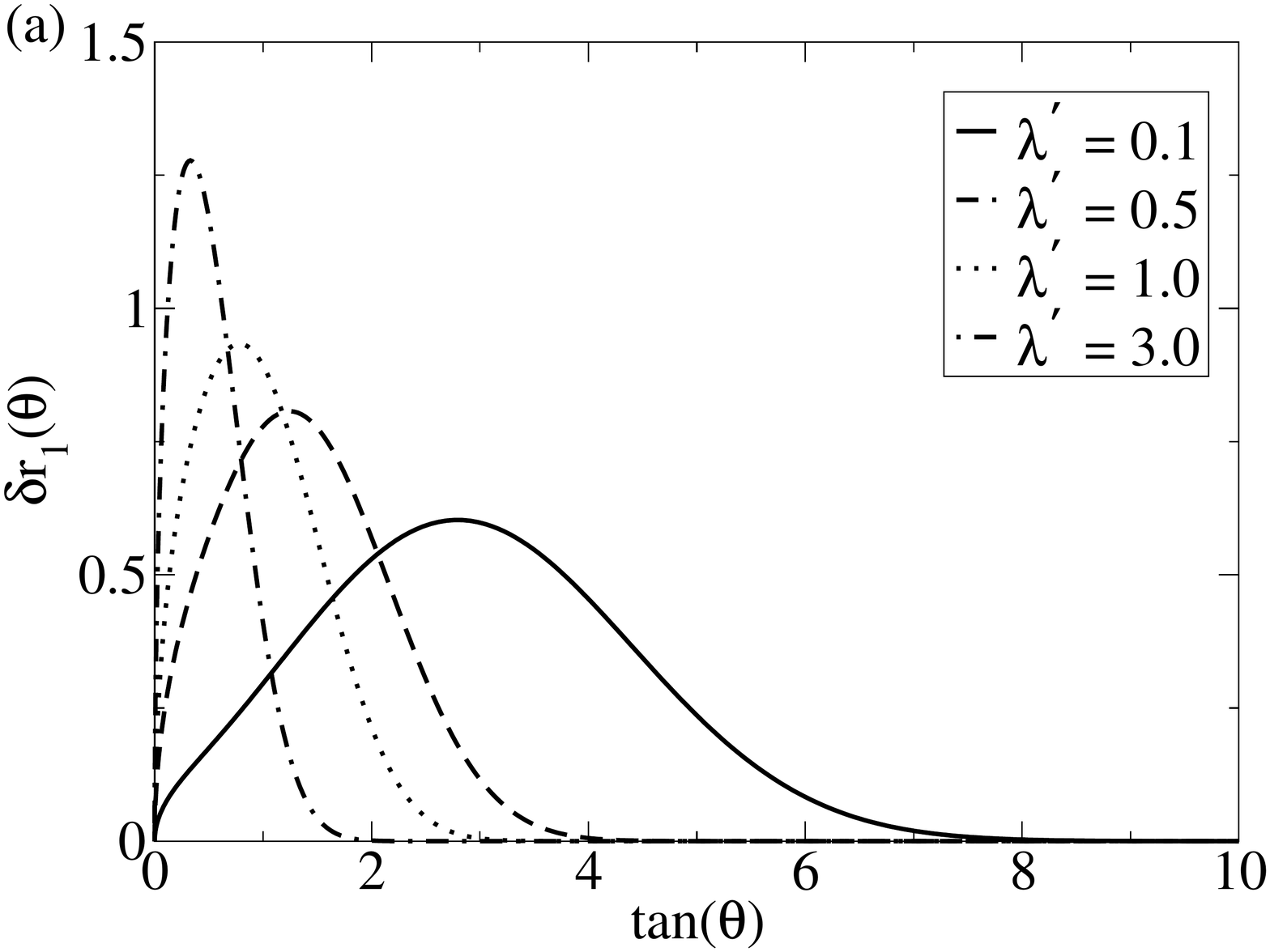}
\includegraphics[width=0.99\columnwidth]{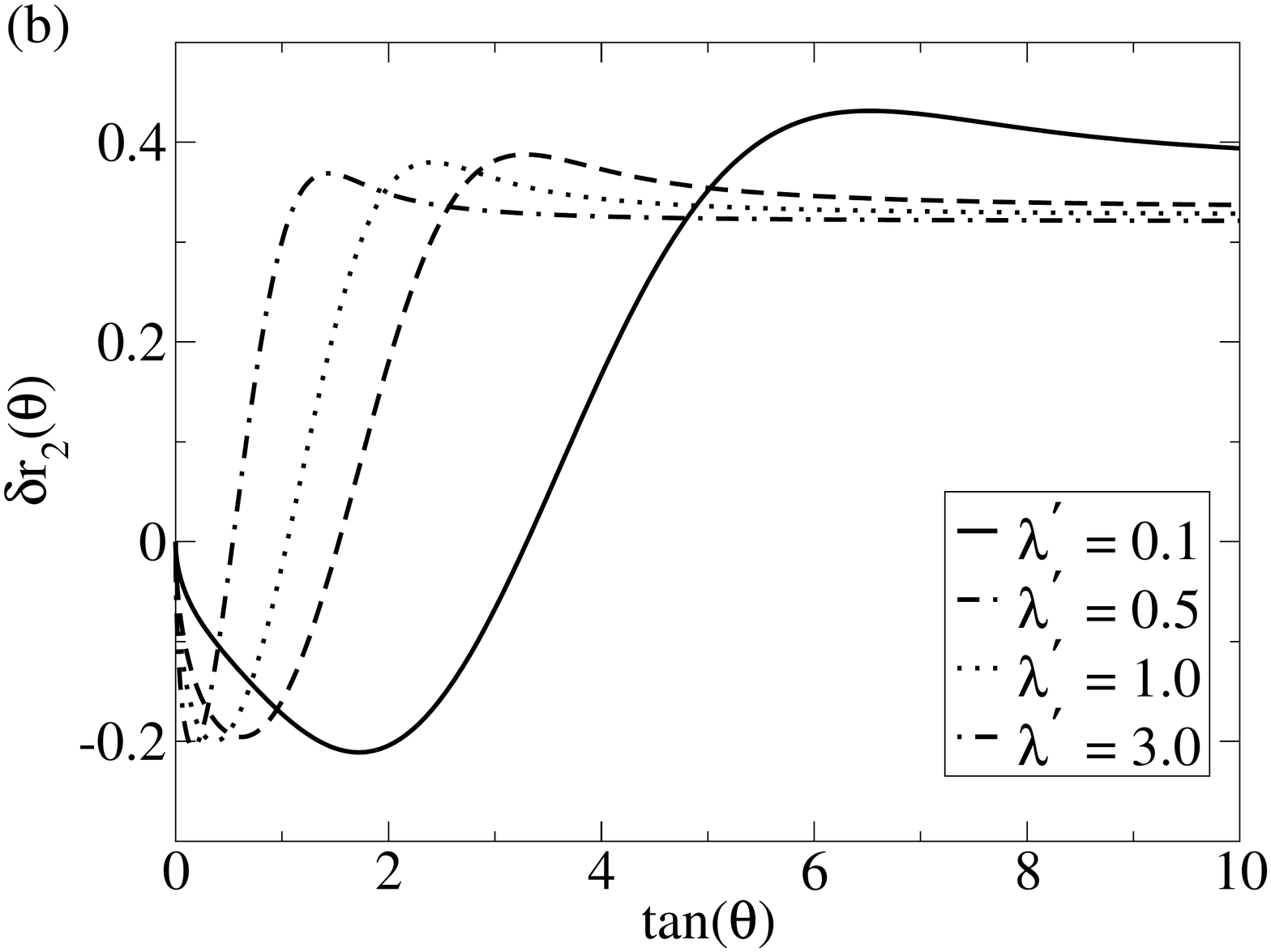}
\caption{The eigenfunctions of Eq. (\ref{dome_ev_eqn}) for $\lambda' = 0.1$, $0.5$, $1.0$ and $3.0$.  (a) The first eigenfunction satisfies the boundary conditions for symmetric modes, implying the instability of the dome solution. (b) The second eigenfunction does not satisfy the boundary condition at infinity.} \label{fig_dome_ef}
\end{figure}

We see from the asymptotic formula, Eq. (\ref{eqn_dome_stability_large_asym2}), that,
\begin{equation}
\delta r_2 (x) \equiv \sqrt{x} g_2(x) \sim 1 + O\left(\frac{1}{x^{5/2}}\right),
\end{equation}
as $x\rightarrow\infty$ or $\theta\rightarrow\pi/2$.  This means that $\delta r_2(\theta)$ does not satisfy the boundary condition, $\delta r(\theta=\pi/2)=0$.  The solution, $\delta r_1(\theta)$, is the only solution that satisfies the boundary conditions, Eq. (\ref{eqn_bc_sym}).

To obtain the full eigenfunctions, we use the asymptotic formula, Eq. (\ref{eqn_dome_stability_large_asym}) and (\ref{eqn_dome_stability_large_asym2}), as initial conditions and integrate numerically from a large value of $x=c$ ($c=10$ in this case) back to $x=0$.  The Gram-Schmidt orthonormalization procedure is employed to ensure the linear independence of the two eigenfunctions.  The eigenfunctions are normalized such that
\begin{equation}
\int_0^c \delta r_i(x)\delta r_j(x) dx = \delta_{ij}.
\end{equation}
Fig. \ref{fig_dome_ef} shows $\delta r_1(\theta)$ and $\delta r_2(\theta)$ for $\lambda' = 0.1$, $0.5$, $1.0$ and $3.0$.  From the graph, we confirm that $\delta r_1(\theta)$ satisfies the boundary conditions, Eq.(\ref{eqn_bc_sym}), while $\delta r_2 (\theta)$ does not.

Note that $\delta r_1(\theta)$ satisfies only the boundary conditions for the symmetric modes, but not the anti-symmetric modes.  We need a linear combination of $\delta r_1(\theta)$ and $\delta r_2(\theta)$ to form a solution that satisfies the latter.  But since $\delta r_2(\theta)$ does not satisfy the boundary condition at $\theta=\pi/2$, such a linear combination would not satisfy it either.

To conclude, there are always solutions to Eq. (\ref{dome_ev_eqn}) satisfying the boundary conditions for the symmetric modes for every positive value of $\lambda$, {\it i.e.}, the domes are unconditionally linearly unstable.  This seems to be a contradiction with the field observation of domes, which are presumably stable.   We will postpone the discussion of this issue to the end of the next section, after we have discussed stalactite formation.

\section{Stalactites}
\label{stalactite}
In studying the formations of travertine domes near geothermal hot
springs, it helps to study a similar geophysical process, namely, the
formation of stalactites, which are cylindrical structures formed by
precipitation of calcium carbonate in limestone caves. Here, we will
summarize the results Short {\it et al.}\cite{SHOR05a,SHOR05b}
obtained and apply our formulation to study the stability of
stalactites.

\subsection{Steady state}

As discussed earlier, the growth rate of stalactites is directly
proportional to the local fluid thickness, $h$.  From the field
observation, stalactite formation shares the following features with
dome formation: They both are circularly symmetrical, formed under a
shallow water laminar flow, and can be assumed to be locally flat.
So, by using the analysis of dome formation, in particular, from Eq.
(\ref{mass_conservation}), we have
\begin{equation}
h = \left( \frac{\beta}{r \sin\theta}\right)^{1/3},
\end{equation}
where $\beta \equiv 3\nu Q/2\pi g$ is a constant. The dynamical equation, Eq. (\ref{growth_eqn}), then becomes
\begin{equation}\label{stal_growth_eqn}
\frac{\partial\kappa}{\partial t}\bigg|_{\theta} = -\kappa^2\left(1+
\frac{\partial^2}{\partial \theta^2}\right)\left[G
\left(\frac{\beta}{r\sin\theta}\right)^{1/3}\right],
\end{equation}
where $G$ depends on water chemistry and the input
flux\cite{SHOR05a,SHOR05b}.  Following the same strategy employed in
the case of travertine domes, we obtain a uniformly translating
solution,
\begin{equation}
r(\theta)= \frac{r_0}{\sin\theta\cos ^3\theta},
\label{stalactite_shape}
\end{equation}
where the tip velocity $v_t$ comes in as an integration constant,
and the scale $r_0\equiv \beta(G/v_t)^3$.  By defining $\rho\equiv
r/r_0$, $z\equiv \zeta/r_0$ and using the trigonometric relation
$\tan\theta=-dz/d\rho$, we obtain
\begin{equation}
\frac{z'}{(1+z')^2}+\frac{1}{\rho}=0,
\end{equation}
which is the result derived in Refs. \cite{SHOR05a,SHOR05b}.

\subsection{Linear stability analysis}

We study the stability of solution Eq. (\ref{stalactite_shape}) by
introducing a perturbation:
\begin{equation}
r(\theta)=\bar{r}(\theta) + \delta r(\theta) e^{\lambda t},
\label{stal_perturb_r}
\end{equation}
where $\bar{r}$ is the unperturbed solution given by Eq.
(\ref{stalactite_shape}) and $\delta r$ is the perturbation.
Substituting Eq. (\ref{stal_perturb_r}) into Eq. (\ref{stal_growth_eqn})
and expanding the resulting equation in $\delta r$ gives
\begin{equation}
\lambda'\frac{d \delta r}{d\theta} + \cos\theta
\left[1+\frac{d^2}{d\theta ^2}\right] \left(\delta r
\sin\theta\cos ^4 \theta \right)=0,
\label{stal_stability_eqn}
\end{equation}
where $\lambda'\equiv 3G^3\lambda/v_t^4$.  We follow the same approach as in the case of the dome and study the asymptotic behaviors of the solutions of Eq. (\ref{stal_stability_eqn}).  For $\theta\rightarrow 0$, we expand Eq. (\ref{stal_stability_eqn}) in $\theta$ and obtain
\begin{equation}
\lambda'\frac{d\delta r}{d\theta}+\left[1+\frac{d^2}{d\theta^2}\right]\theta\delta r=0,
\end{equation}
whose solution is given by $r\sim\theta^{\sigma}$, where $\sigma = -1-\lambda$.  Because
$\sigma<0$ for all $\lambda>0$, the solution diverges as $\theta\rightarrow 0$.  This shows that there are no eigenmodes for $\lambda>0$.  As a result, we conclude
that the steady-state solution Eq. (\ref{stalactite_shape}) is linearly
stable against the class of perturbations considered here.

Let us also look at the asymptotics as $x \rightarrow \infty$ for completeness.
Following the strategy employed in the study of dome stability, we make the transformation
$g(\theta)=\tan\theta \delta r(\theta)$ and $x=\tan\theta$.  Eq. (\ref{stal_stability_eqn})
then becomes
\begin{equation}
\frac{d^2 g}{dx^2} + u(x)\frac{dg}{dx} +v(x)g(x) = 0,
\end{equation}
where
\begin{equation}
u(x) = \frac{-8 x}{1+x^2} + \frac{\lambda' (1+x^2)^{3/2}}{x},
\end{equation}
and,
\begin{equation}
v(x) = \frac{\lambda'(1+x^2)^{3/2}}{x^2} +\frac{20x^2 -5}{(1+x^2)^2} + \frac{1}{(1+x^2)^{5/2}}.
\end{equation}
As $x\rightarrow\infty$,
\begin{equation}
u(x) \sim \lambda' x^2 + \frac{3\lambda'}{2} - \frac{8}{x} + \frac{3\lambda'}{8x^2}+\frac{8}{x^3}+O\left(\frac{1}{x^4}\right),
\end{equation}
and,
\begin{equation}
v(x) \sim \lambda' x + \frac{3\lambda'}{2 x}+ \frac{20}{x^2}++ \frac{3\lambda'}{8x^3}+O\left(\frac{1}{x^4}\right).
\end{equation}
By following the same asymptotic analysis as we did in the last section, we get,
\begin{equation}
g_1(x) \sim \exp\left(\frac{-\lambda' x^3}{3}-\frac{3\lambda'x}{2}\right),
\end{equation}
and
\begin{equation}
g_2(x) \sim \frac{1}{x} + \frac{10}{\lambda'x^4} - \frac{98}{5\lambda'x^6} + O\left(\frac{1}{x^7}\right).
\end{equation}

\begin{figure}[t]
\includegraphics[width=0.99\columnwidth]{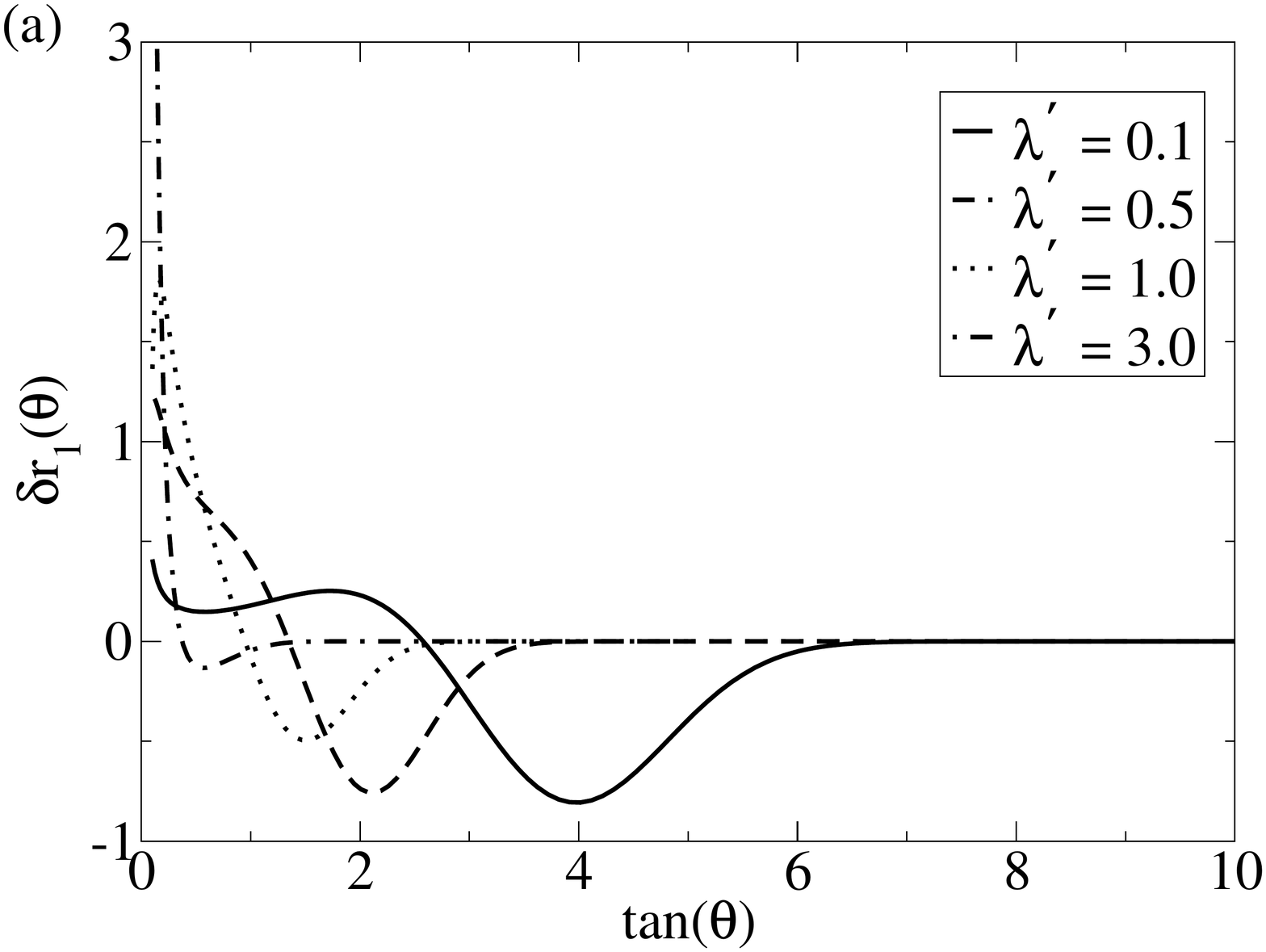}
\includegraphics[width=0.99\columnwidth]{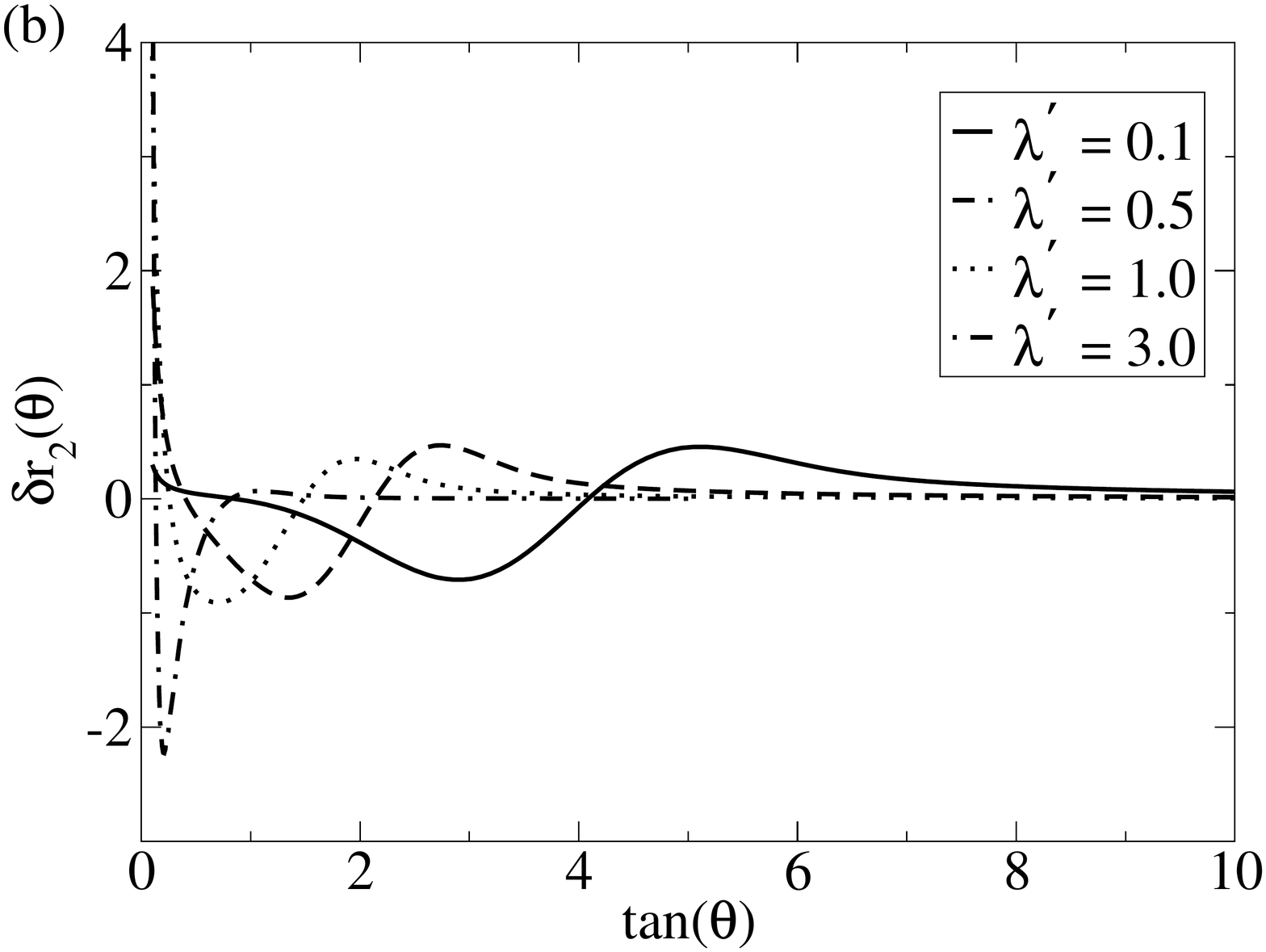}
\caption{The eigenfunctions of Eq. (\ref{stal_stability_eqn}) for $\lambda' = 0.1$, $0.5$, $1.0$ and $3.0$.  These solutions do not satisfy the boundary conditions, as they all diverges at $\theta=0$.} \label{fig_sta_ef}
\end{figure}

These can be used as the initial conditions to integrate numerically from a large value of $x$, giving the full eigenfunctions.  Again, the Gram-Schmidt orthonormalization procedure is employed.  The two branches of solutions, $\delta r_{1,2}(\theta)$, are plotted in Fig. \ref{fig_sta_ef}.  They do not satisfy the boundary conditions as they both diverge at $\theta = 0$.  So the stalactite solution is stable.

\section{Comparison between domes and stalactites}
\label{domes-stalactites-discussion}

We have shown that there is a continuous spectrum of unstable modes for
travertine domes, but stalactites, which are formed by an apparently
similar process, are predicted to be linearly stable.  We need to (a)
explain why it is that domes can be observed in the field, and (b)
interpret the source of the difference in stability between the two
seemingly-related growth motifs.  We initially found it surprising that
there is a qualitative difference in stability, even though the
dynamics of domes and stalactites seem to differ in only relatively
minor ways: the growth of domes depends on the depth-averaged fluid
velocity whereas the growth of stalactites depends on the fluid
thickness.  In both cases, the approximation of local flatness is used,
so this is unlikely to be the source of the difference.

Our interpretation is that the difference in stability arises from the
direction of growth, and as a result, the manner in which surface
tension effects correct the zeroth order solutions we have discussed.
The direction of growth is important, because it dictates the way in
which shape perturbations propagate.  For domes growing with
sufficiently large $v_t$, shape perturbations are advected away from
the vent down the body of the dome, in a manner reminiscent of the way
in which shape perturbations are advected down the body of a growing
dendrite\cite{barber1987dds}.  These perturbations may also grow during
this process, but the development of this instability is in practice
regularized by any non-zero surface tension, leading to contact line
formation, film break-up and the formation of rivulets. This heuristic
argument is supported by the shape of the linear stability
eigenfunctions shown in Fig. \ref{fig_dome_ef}.  For
stalactites, on the other hand, the fluid becomes increasingly thick as
it flows down towards the tip, and perturbations only increase the
growth velocity of the tip, rather than cause growing instabilities
away from the tip.  Thus, the only place where surface tension is
significant is at the tip of the stalactite, where the surface tension
holds a water droplet until the droplet becomes too heavy and drops.
This dynamics, we believe, mainly contributes to the precipitation rate
at the tip, which affects only the growth rate of the whole stalactite.
In other words, it only renormalizes the value of $v_t$, which, in any
case, is a fitting parameter. Surface tension is, therefore, not
important in the dynamics of stalactite formation and it should not
affect its stability.

Returning now to the case of travertine domes, we conclude that the unstable modes are small near the vent and grow in amplitude near the tail of the dome.  This, however, is precisely the region where the film becomes thin and contact line formation can occur, leading to the fluting pattern observed in the real systems.  The precipitation rate in this region is also lower, due to the depleted Ca$^{2+}$ concentration, and this helps stabilizing the domes too.  It is possible that the growth of the instabilities predicted here triggers the formation of contact lines and film break-up.  Thus, we conclude that the dome is in some sense similar to the problem of dendritic growth, where a smooth tip is followed by a train of sidebranches, widely interpreted to be
due to a noise-induced instability\cite{pieters1986nds, GONZ01}.  It is possible that the full inclusion of surface tension in the model would have as important a role in selection and stability as it does in dendritic growth\cite{benjacob1984psd, kessler1985gmi}.

\section{Damming instability}\label{sec_damming}
Having studied the formation of domes and stalactites, we now try to
understand some aspects of the large scale morphology of hot spring
landscapes.  We see in Fig. \ref{AT3}b that the pattern formed is
complicated, with ponds of similar shapes but different sizes.
Empirical data shows that the distribution of pond sizes indeed follows
a power law\cite{VEYS06}. This scale-invariant pattern hints at an
underlying scale-invariant precipitation dynamics, {{\it i.e.}}, a
dynamics without a characteristic length scale.

It is difficult to predict analytically the statistical properties of
the landscape, such as the pond size distribution, due to the
mathematical complexity of the equations involved. We can,
nevertheless, study a simple case of precipitation over a planar slope.
By studying the linear stability of this dynamics, we should be able to
expose the essential physics of the formation of these scale-invariant
patterns.  The nonlinear regime of the modeling can be studied using
the cellular model we introduced earlier.  In this section, we consider
a one-dimensional flow down an inclined plane, and evaluate the linear
stability spectrum.

The fluid flow in travertine systems is, unlike in the cases of dome
and stalactite formations, generally turbulent.  It is therefore
necessary to use the formulation of Eq. (\ref{Dressler_eqn_1})-(\ref{Dressler_eqn_2}).  The
turbulent drag leads to a steady flow regime, about which we linearize.
Since the angle $\theta$ is the same along a constant slope, it is more
convenient to use the arc length, $s$, as the independent variable in
the growth equation, so the dynamics of local curvature, $\kappa$, is
given by
\begin{equation}
\frac{\partial\kappa}{\partial t}\bigg|_{n} = -\left(\kappa^2+
\frac{\partial^2}{\partial s^2}\right)G u_0,
\end{equation}
where the subscript $n$ denotes a derivative taken at a point moving along the outward normal of the curve.  This, together with the Dressler equation, Eq. (\ref{Dressler_eqn_1})-(\ref{Dressler_eqn_2}),
gives the complete description of the system.

We scale the independent variables to their natural units,
\begin{equation}
t' = \frac{U}{R}t,\quad s' = \frac{s}{R},\quad \zeta ' = \frac{\zeta}{R},
\end{equation}
and define the following dimensionless variables,
\begin{equation}
u_0' = \frac{u_0}{U},\quad h' = \frac{h}{H},\quad\kappa ' = R\kappa,\quad\sigma \equiv \frac{H}{R},
\end{equation}
where $U$, $H$ and $R$ are the characteristic scales of the fluid velocity, fluid thickness and the landscape respectively, and $\sigma$ is the ratio between the $H$ and $R$, which is small in the regime of shallow water flow.  The governing equations then become (we drop all the primes on the variables for simplicity),
\begin{equation}
\left.\frac{\partial \kappa}{\partial t}\right|_n = -\left(\kappa^2 + \frac{\partial^2}{\partial s^2}\right)G u_0,
\end{equation}
\begin{equation}
\sigma F_m\frac{\partial u_0}{\partial t} + \frac{\partial E}{\partial s} = \frac{-C_fF_mu_0^2}{h\left(1-\frac{\sigma\kappa h}{2}\right)},
\end{equation}
\begin{equation}
(1-\sigma\kappa h)\sigma\frac{\partial h}{\partial t} + \frac{\partial q}{\partial s} = 0,
\end{equation}
with
\begin{equation}
E = \zeta + \sigma h \cos\theta + \frac{p_h}{\rho g} + \frac{\sigma F_m u_0^2}{2(1-\sigma\kappa h)^2},
\end{equation}
\begin{equation}
q = \frac{-u_0}{\kappa}\ln(1-\sigma\kappa h),
\end{equation}
where we defined the Froude number, $F_m\equiv U^2/gR$.

The uniform solution of this set of equations is given by
\begin{equation}
\bar{u}_0 = \sqrt{\frac{\sin\theta}{C_fF_m}},
\end{equation}
\begin{equation}
\bar{h} = 1,
\end{equation}
\begin{equation}
\bar{\theta} = \theta_0
\end{equation}
\begin{equation}
\bar{\kappa} = 0,
\end{equation}
where $\theta_0$ is the initial inclination of the slope. The
linear stability analysis is performed by adding harmonic perturbations
to the solution,
\begin{equation}
u_0 = \bar{u}_0 + \delta u_0 e^{ips + \lambda t},
\end{equation}
\begin{equation}
h = 1 + \delta h e^{ips + \lambda t},
\end{equation}
\begin{equation}
\theta = \bar{\theta} + \delta\theta e^{ips + \lambda t},
\end{equation}
\begin{equation}
\kappa \equiv \frac{\partial \theta}{\partial s} = ip\delta\theta e^{ips + \lambda t},
\end{equation}
and linearizing the resultant equations to the first order in the
perturbations, resulting in three equations for $\delta u_0$, $\delta h$ and $\delta \theta$,
\begin{equation}
ip\lambda\delta\theta = p^2G\delta u_0
\end{equation}
\begin{equation}
(\lambda+ip\bar{u_0})\delta h + ip\delta u_0 - \frac{\sigma u_0p^2}{2}\delta\theta = 0,
\end{equation}
\begin{eqnarray}
\sigma F_m\lambda \delta u_0
&=& \delta\theta (\cos\bar{\theta} + ip\sigma\sin\bar{\theta} + p^2\sigma^2\bar{u}_0^2F_m)\nonumber\\
&&
 - \delta\theta\frac{C_fF_m\bar{u}_0^2\sigma ip}{2}\nonumber\\
&&
+ \delta h (-ip\sigma\cos\bar{\theta} + C_fF_m\bar{u}_0^2)\nonumber\\
&&
+ \delta u_0 (-ip \sigma F_m u_0 - 2C_fF_m\bar{u}_0)
\end{eqnarray}

\begin{figure}
\includegraphics[width=0.49\columnwidth]{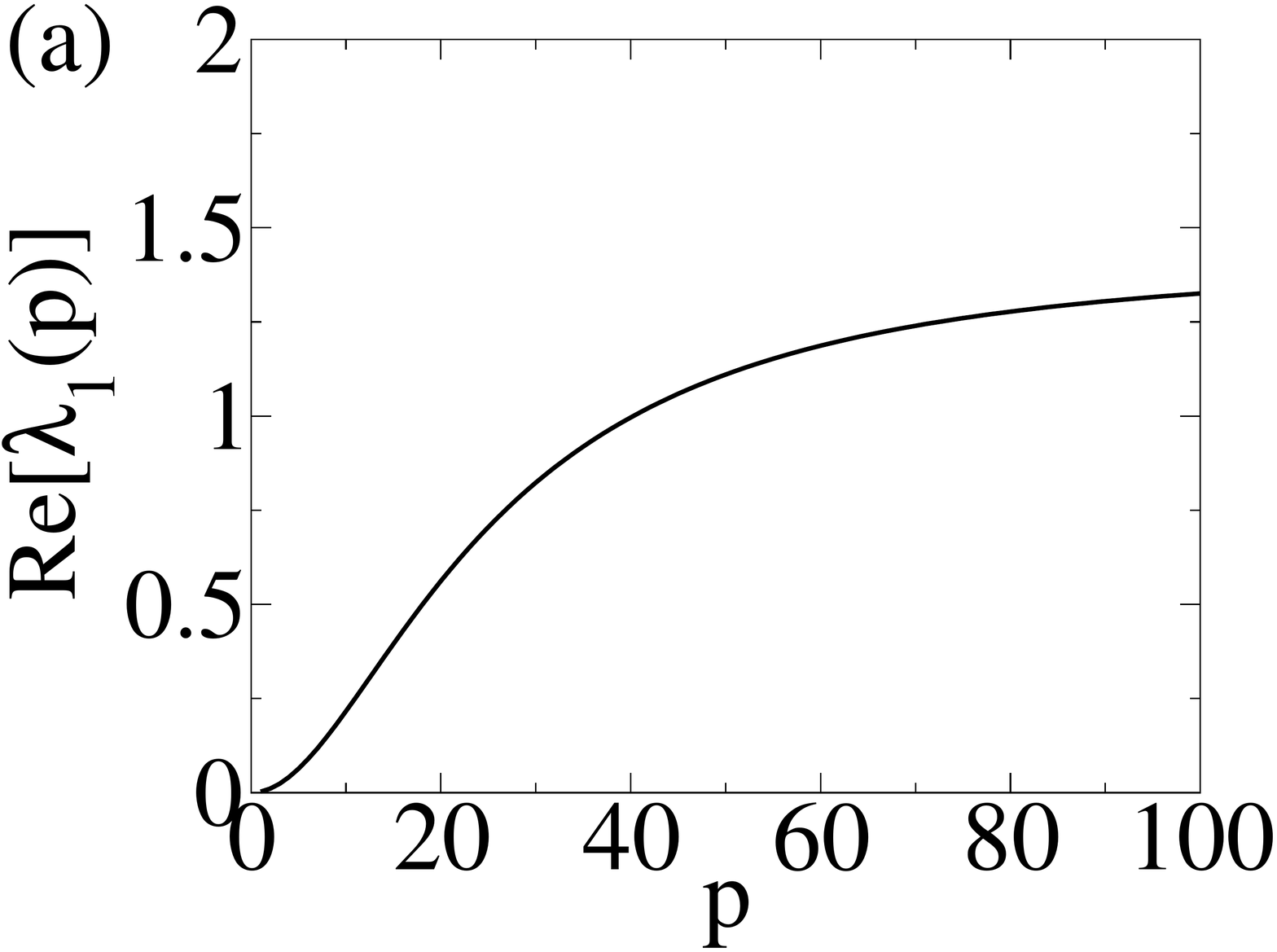}
\includegraphics[width=0.49\columnwidth]{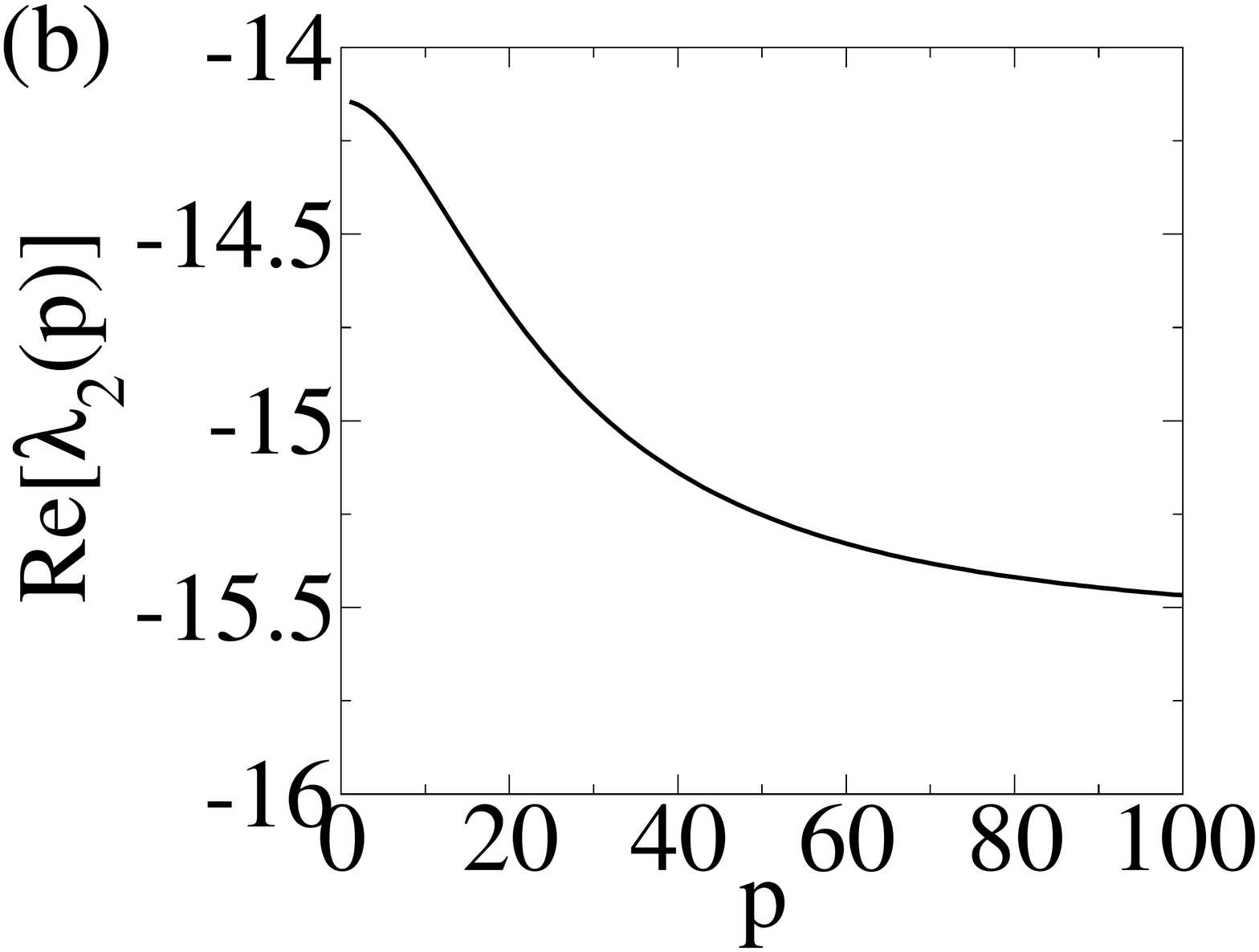}
\includegraphics[width=0.49\columnwidth]{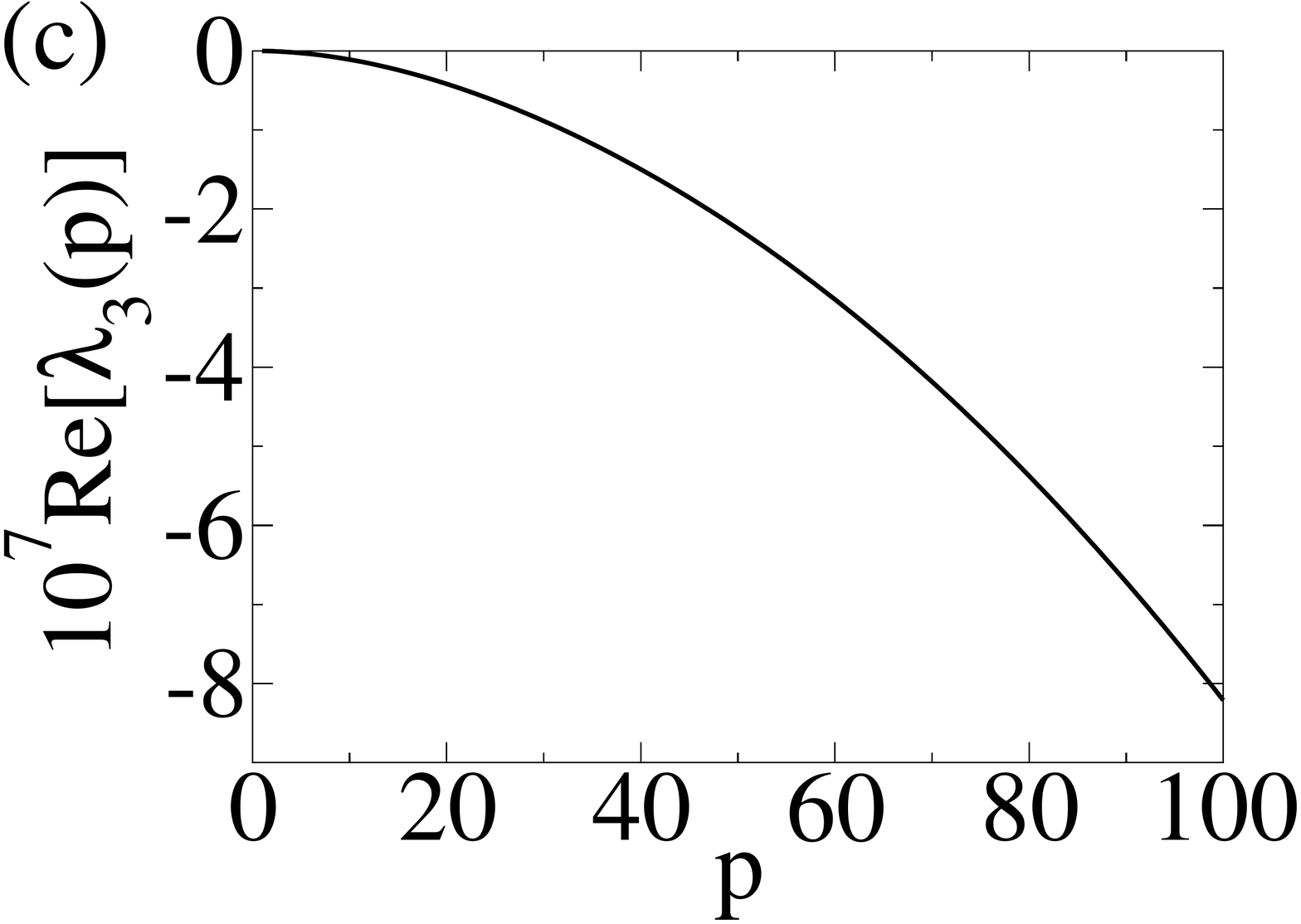}
\includegraphics[width=0.49\columnwidth]{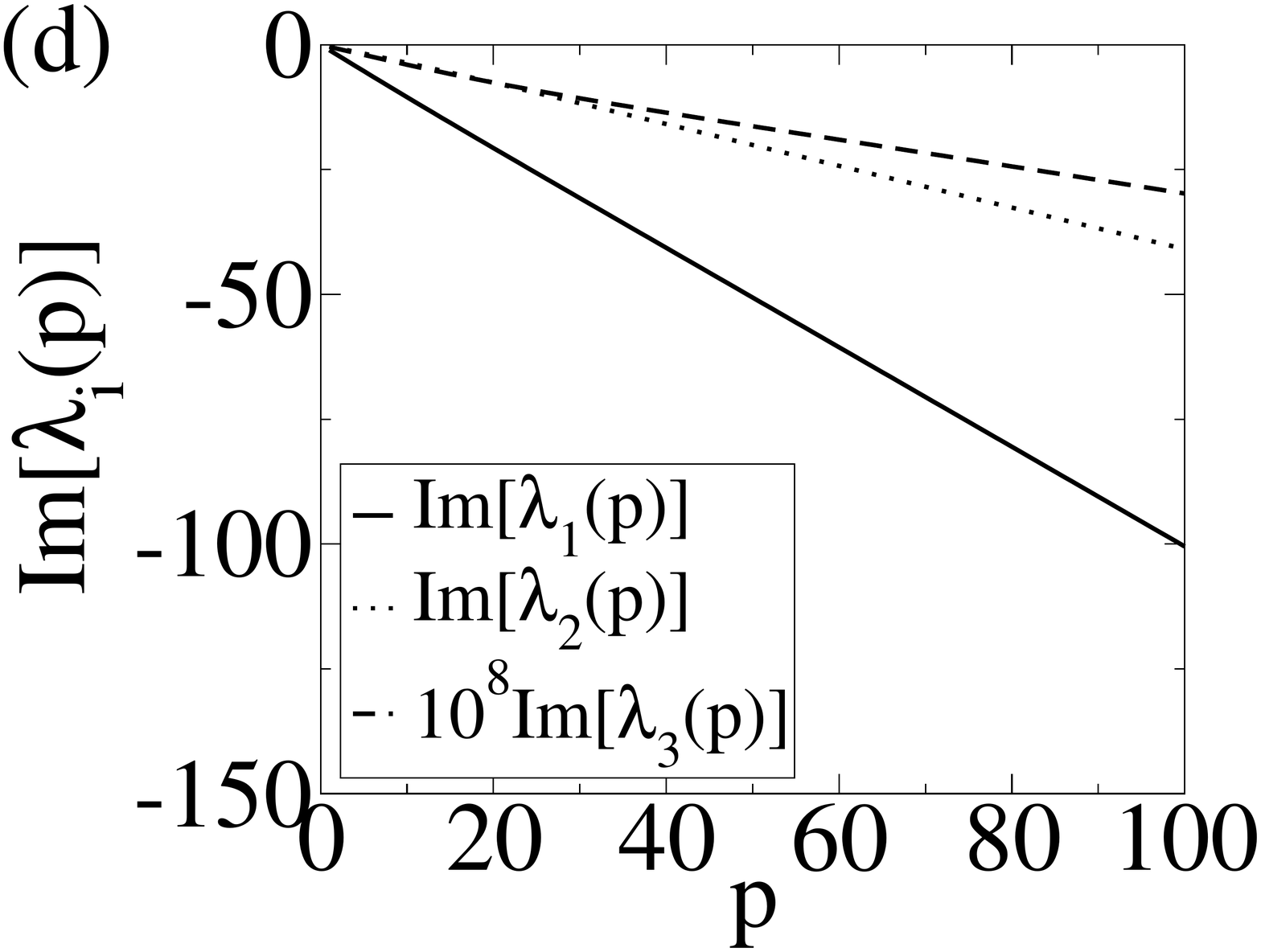}
\caption{The damming instability spectrum with parameters $(\theta_0, G, F_m, C_f, \sigma) = (\pi/6, 10^{-8}, 10, 0.1, 0.01)$.  (a)-(c) The real parts of the three branches of solutions.  The first branch, $\lambda_1$, is positive for all $p$, implying that the solution is unconditionally linearly unstable. (d) The imaginary parts of the solutions.}
\label{fig_damming_spectrum}
\end{figure}

A single dispersion relation can be obtained by combining all three
equations and eliminating $\delta u_0$, $\delta h$ and
$\delta\theta$.  The result is a cubic equation in $\lambda$,
\begin{equation}
\lambda^3 + a_2(p)\lambda^2 + a_1(p)\lambda + a_0(p) = 0,
\label{eqn_dispersion}
\end{equation}
where
\begin{equation}
a_2(p) = 2i\bar{u}_0 p + \frac{2 C_f \bar{u}_0}{\sigma},
\end{equation}
\begin{eqnarray}
a_1(p) &=& p^3 i\sigma\bar{u}_0^2G\nonumber\\
&&
+ p^2 \left(\frac{G\sin\theta}{F_m} + \frac{C_f \bar{u}_0^2G}{2} + \frac{\cos\bar{\theta}}{F_m} - \bar{u}_0^2 \right)\nonumber\\
&&
+ p \left( \frac{iG\cos\bar{\theta}}{\sigma F_m} + \frac{3iC_f \bar{u}_0^2}{\sigma} \right),
\end{eqnarray}
\begin{eqnarray}
a_0(p) &=& p^4 \left( -\sigma\bar{u}_0^3 G + \frac{\sigma\bar{u}_0G\cos\bar{\theta}}{2F_m} \right)\nonumber\\
&&
+ p^3 \left(\frac{-i G\bar{u}_0}\sin\bar{\theta}{F_m} + iGC_f\bar{u}_0^3\right)\nonumber\\
&&
+ p^2 \left( \frac{-G\bar{u}_0\cos\bar{\theta}}{\sigma F_m} \right).
\end{eqnarray}

For the parameter set $(\theta_0, G, F_m, C_f, \sigma) = (\pi/6, 10^{-8}, 10, 0.1, 0.01)$, the three roots of the Eq. (\ref{eqn_dispersion}), $\lambda_i$, are computed numerically and are plotted in Fig. \ref{fig_damming_spectrum}.  From the graph, we see the first branch of the solutions is always unstable, while the remaining two branches are always stable, implying that the solution is unconditionally linearly unstable.  This is the damming instability.

To conclude, we found that the trivial flow down a constant inclined plane is unstable towards all length scales, suggesting that when fully developed into the nonlinear regime, the landscape would have no selected length scale - a surmise in accord with field observations and our cell dynamical system simulations.

\section{Conclusion}\label{sec_conclusion}
By combining fluid dynamics and surface growth kinematics, we
formulated a mathematical framework to study geological pattern
formation due to carbonate precipitation and applied it to study the
formation and stability of a variety of motifs.  The theory
successfully predicted the shape of observed spherically symmetric
domes for angle $\theta$ less than a critical angle $\theta _c$.  By
comparing with results from a cellular model, we showed that the
departure of our theoretical prediction from observation for $\theta >
\theta_c$ is due to the neglect of surface tension.  We also showed
that domes are linearly unstable towards axisymmetric perturbations,
but the instability is manifested in the tail of the dome away from the
vent.  The instability is masked by the thinning of the fluid film and
ultimately the formation of contact lines due to surface tension. This
contrasted with the case of stalactites, whose growth forms are
linearly stable to axisymmetric perturbations. The difference between
the stabilities of the dome and stalactite solutions is attributed to
the different geometries and the different role surface tension plays
in these two systems.

This formulation cannot predict the complex landscape formed in the
fully nonlinear regime, but a linear stability analysis for a
one-dimensional flow showed that the apparent scale-invariant landscape
is consistent with our equations.  In future work, we hope to examine
the full two-dimensional instability problem, in order to investigate
the dynamics of pond formation, possibly as a transverse morphological
instability, akin to meandering in step-flow processes on vicinal
surfaces\cite{bales1990mit}.

\begin{acknowledgments}

We acknowledge stimulating discussions with all the members of the
University of Illinois Yellowstone Group, but especially Bruce Fouke,
Michael Kandianis and John Veysey, whose expertise, collaboration and
review of this manuscript we have greatly enjoyed and appreciated. This
material is based upon work supported by the National Science
Foundation under Grant No. NSF-EAR-0221743. Any opinions, findings, and
conclusions or recommendations expressed in this material are those of
the authors and do not necessarily reflect the views of the National
Science Foundation.

\end{acknowledgments}

\bibliographystyle{apsrev}

\bibliography{geophy_pattern_bib}

\begin{thebibliography}{40}
\expandafter\ifx\csname natexlab\endcsname\relax\def\natexlab#1{#1}\fi
\expandafter\ifx\csname bibnamefont\endcsname\relax
  \def\bibnamefont#1{#1}\fi
\expandafter\ifx\csname bibfnamefont\endcsname\relax
  \def\bibfnamefont#1{#1}\fi
\expandafter\ifx\csname citenamefont\endcsname\relax
  \def\citenamefont#1{#1}\fi
\expandafter\ifx\csname url\endcsname\relax
  \def\url#1{\texttt{#1}}\fi
\expandafter\ifx\csname urlprefix\endcsname\relax\def\urlprefix{URL }\fi
\providecommand{\bibinfo}[2]{#2}
\providecommand{\eprint}[2][]{\url{#2}}

\bibitem[{\citenamefont{Wooding}(1991)}]{WOOD91}
\bibinfo{author}{\bibfnamefont{R.~A.} \bibnamefont{Wooding}},
  \bibinfo{journal}{J. Geophysical Res.} \textbf{\bibinfo{volume}{96}},
  \bibinfo{pages}{667} (\bibinfo{year}{1991}).

\bibitem[{\citenamefont{Goldenfeld et~al.}(2006)\citenamefont{Goldenfeld, Chan,
  and Veysey}}]{GOLD06}
\bibinfo{author}{\bibfnamefont{N.}~\bibnamefont{Goldenfeld}},
  \bibinfo{author}{\bibfnamefont{P.~Y.} \bibnamefont{Chan}}, \bibnamefont{and}
  \bibinfo{author}{\bibfnamefont{J.}~\bibnamefont{Veysey}},
  \bibinfo{journal}{Physical Review Letters} \textbf{\bibinfo{volume}{96}},
  \bibinfo{pages}{254501} (\bibinfo{year}{2006}).

\bibitem[{\citenamefont{Fouke et~al.}(2000)\citenamefont{Fouke, Farmer, Marais,
  Pratt, Sturchio, and Discipulo}}]{FOUK00}
\bibinfo{author}{\bibfnamefont{B.~W.} \bibnamefont{Fouke}},
  \bibinfo{author}{\bibfnamefont{J.~D.} \bibnamefont{Farmer}},
  \bibinfo{author}{\bibfnamefont{D.~J.~D.} \bibnamefont{Marais}},
  \bibinfo{author}{\bibfnamefont{L.}~\bibnamefont{Pratt}},
  \bibinfo{author}{\bibfnamefont{N.~C.} \bibnamefont{Sturchio}},
  \bibnamefont{and} \bibinfo{author}{\bibfnamefont{M.~K.}
  \bibnamefont{Discipulo}}, \bibinfo{journal}{J. Sed. Res.}
  \textbf{\bibinfo{volume}{70}}, \bibinfo{pages}{265} (\bibinfo{year}{2000}).

\bibitem[{\citenamefont{Fouke}(2001)}]{FOUK01}
\bibinfo{author}{\bibfnamefont{B.~W.} \bibnamefont{Fouke}},
  \bibinfo{journal}{J. Sed. Res.} \textbf{\bibinfo{volume}{71}},
  \bibinfo{pages}{497} (\bibinfo{year}{2001}).

\bibitem[{\citenamefont{Hammer et~al.}(2006)\citenamefont{Hammer, Dysthe, and
  Jamtveit}}]{HAMM06}
\bibinfo{author}{\bibfnamefont{O.}~\bibnamefont{Hammer}},
  \bibinfo{author}{\bibfnamefont{D.}~\bibnamefont{Dysthe}}, \bibnamefont{and}
  \bibinfo{author}{\bibfnamefont{B.}~\bibnamefont{Jamtveit}},
  \bibinfo{journal}{Arxiv preprint physics/0601116}  (\bibinfo{year}{2006}).

\bibitem[{\citenamefont{Short et~al.}(2005{\natexlab{a}})\citenamefont{Short,
  Baygents, Beck, Stone, III, and Goldstein}}]{SHOR05a}
\bibinfo{author}{\bibfnamefont{M.~B.} \bibnamefont{Short}},
  \bibinfo{author}{\bibfnamefont{J.~C.} \bibnamefont{Baygents}},
  \bibinfo{author}{\bibfnamefont{J.~W.} \bibnamefont{Beck}},
  \bibinfo{author}{\bibfnamefont{D.~A.} \bibnamefont{Stone}},
  \bibinfo{author}{\bibfnamefont{R.~S.~T.} \bibnamefont{III}},
  \bibnamefont{and} \bibinfo{author}{\bibfnamefont{R.~E.}
  \bibnamefont{Goldstein}}, \bibinfo{journal}{Phys. Rev. Lett.}
  \textbf{\bibinfo{volume}{94}}, \bibinfo{pages}{018501}
  (\bibinfo{year}{2005}{\natexlab{a}}).

\bibitem[{\citenamefont{Short et~al.}(2005{\natexlab{b}})\citenamefont{Short,
  Baygents, and Goldstein}}]{SHOR05b}
\bibinfo{author}{\bibfnamefont{M.~B.} \bibnamefont{Short}},
  \bibinfo{author}{\bibfnamefont{J.~C.} \bibnamefont{Baygents}},
  \bibnamefont{and} \bibinfo{author}{\bibfnamefont{R.~E.}
  \bibnamefont{Goldstein}}, \bibinfo{journal}{Phys. Fluids}
  \textbf{\bibinfo{volume}{17}}, \bibinfo{pages}{083101}
  (\bibinfo{year}{2005}{\natexlab{b}}).

\bibitem[{\citenamefont{Pye and Tsoar}(1990)}]{PYE90}
\bibinfo{author}{\bibfnamefont{K.}~\bibnamefont{Pye}} \bibnamefont{and}
  \bibinfo{author}{\bibfnamefont{H.}~\bibnamefont{Tsoar}},
  \emph{\bibinfo{title}{{Aeolian sand and sand dunes}}}
  (\bibinfo{publisher}{Unwin Hyman Boston}, \bibinfo{year}{1990}).

\bibitem[{\citenamefont{Lancaster}(1996)}]{LANC96}
\bibinfo{author}{\bibfnamefont{N.}~\bibnamefont{Lancaster}},
  \emph{\bibinfo{title}{{Geomorphology of Desert Dunes}}}
  (\bibinfo{publisher}{Routledge}, \bibinfo{year}{1996}).

\bibitem[{\citenamefont{Kerr and Turner}(1996)}]{kerr1996cag}
\bibinfo{author}{\bibfnamefont{R.}~\bibnamefont{Kerr}} \bibnamefont{and}
  \bibinfo{author}{\bibfnamefont{J.}~\bibnamefont{Turner}},
  \bibinfo{journal}{Journal of Geophysical Research}
  \textbf{\bibinfo{volume}{101}}, \bibinfo{pages}{25} (\bibinfo{year}{1996}).

\bibitem[{\citenamefont{Goehring et~al.}(2006)\citenamefont{Goehring, Morris,
  and Lin}}]{goehring2006eis}
\bibinfo{author}{\bibfnamefont{L.}~\bibnamefont{Goehring}},
  \bibinfo{author}{\bibfnamefont{S.}~\bibnamefont{Morris}}, \bibnamefont{and}
  \bibinfo{author}{\bibfnamefont{Z.}~\bibnamefont{Lin}},
  \bibinfo{journal}{Physical Review E} \textbf{\bibinfo{volume}{74}},
  \bibinfo{pages}{36115} (\bibinfo{year}{2006}).

\bibitem[{\citenamefont{Murray and Paola}(1994)}]{murray1994cmb}
\bibinfo{author}{\bibfnamefont{A.}~\bibnamefont{Murray}} \bibnamefont{and}
  \bibinfo{author}{\bibfnamefont{C.}~\bibnamefont{Paola}},
  \bibinfo{journal}{Nature} \textbf{\bibinfo{volume}{371}}, \bibinfo{pages}{54}
  (\bibinfo{year}{1994}).

\bibitem[{\citenamefont{Herman and Lorah}(1987)}]{HERM87}
\bibinfo{author}{\bibfnamefont{J.~S.} \bibnamefont{Herman}} \bibnamefont{and}
  \bibinfo{author}{\bibfnamefont{M.~M.} \bibnamefont{Lorah}},
  \bibinfo{journal}{Chemical Geology} \textbf{\bibinfo{volume}{62}},
  \bibinfo{pages}{251} (\bibinfo{year}{1987}).

\bibitem[{\citenamefont{Zhang et~al.}(2001)\citenamefont{Zhang, Zhang, Zhu, and
  Cheng}}]{ZHAN01}
\bibinfo{author}{\bibfnamefont{D.~D.} \bibnamefont{Zhang}},
  \bibinfo{author}{\bibfnamefont{Y.}~\bibnamefont{Zhang}},
  \bibinfo{author}{\bibfnamefont{A.}~\bibnamefont{Zhu}}, \bibnamefont{and}
  \bibinfo{author}{\bibfnamefont{X.}~\bibnamefont{Cheng}}, \bibinfo{journal}{J.
  Sed. Res.} \textbf{\bibinfo{volume}{71}}, \bibinfo{pages}{205}
  (\bibinfo{year}{2001}).

\bibitem[{\citenamefont{Barnes and O'Neil}(1971)}]{BARN71}
\bibinfo{author}{\bibfnamefont{I.}~\bibnamefont{Barnes}} \bibnamefont{and}
  \bibinfo{author}{\bibfnamefont{J.~R.} \bibnamefont{O'Neil}},
  \bibinfo{journal}{Geochimica et Cosmochimica Acta}
  \textbf{\bibinfo{volume}{35}}, \bibinfo{pages}{699} (\bibinfo{year}{1971}).

\bibitem[{\citenamefont{Chafetz et~al.}(1991)\citenamefont{Chafetz, Rush, and
  Utech}}]{CHAF91}
\bibinfo{author}{\bibfnamefont{H.~S.} \bibnamefont{Chafetz}},
  \bibinfo{author}{\bibfnamefont{P.~F.} \bibnamefont{Rush}}, \bibnamefont{and}
  \bibinfo{author}{\bibfnamefont{N.~M.} \bibnamefont{Utech}},
  \bibinfo{journal}{Sedimentology} \textbf{\bibinfo{volume}{38}},
  \bibinfo{pages}{107} (\bibinfo{year}{1991}).

\bibitem[{\citenamefont{Busenberg and Plummer}(1986)}]{BUSE86}
\bibinfo{author}{\bibfnamefont{E.}~\bibnamefont{Busenberg}} \bibnamefont{and}
  \bibinfo{author}{\bibfnamefont{L.~N.} \bibnamefont{Plummer}}, in
  \emph{\bibinfo{booktitle}{US Geological Survey, Bulletin 1578}}, edited by
  \bibinfo{editor}{\bibfnamefont{F.~A.} \bibnamefont{Mumpton}}
  (\bibinfo{year}{1986}), pp. \bibinfo{pages}{139--168}.

\bibitem[{\citenamefont{Renaut and Jones}(1996)}]{RENA96}
\bibinfo{author}{\bibfnamefont{R.~W.} \bibnamefont{Renaut}} \bibnamefont{and}
  \bibinfo{author}{\bibfnamefont{B.}~\bibnamefont{Jones}},
  \bibinfo{journal}{Canadian Journal of Earth Sciences}
  \textbf{\bibinfo{volume}{34}}, \bibinfo{pages}{801} (\bibinfo{year}{1996}).

\bibitem[{\citenamefont{Veysey and Goldenfeld}()}]{VEYS07}
\bibinfo{author}{\bibfnamefont{J.}~\bibnamefont{Veysey}} \bibnamefont{and}
  \bibinfo{author}{\bibfnamefont{N.}~\bibnamefont{Goldenfeld}},
  \bibinfo{note}{unpublished}.

\bibitem[{\citenamefont{Veysey}(2006)}]{VEYS06}
\bibinfo{author}{\bibfnamefont{J.}~\bibnamefont{Veysey}}, Ph.D. thesis,
  \bibinfo{school}{University of Illinois at Urbana-Champaign}
  (\bibinfo{year}{2006}).

\bibitem[{\citenamefont{Brower et~al.}(1983)\citenamefont{Brower, Kessler,
  Koplik, and Levine}}]{brower1983gam}
\bibinfo{author}{\bibfnamefont{R.}~\bibnamefont{Brower}},
  \bibinfo{author}{\bibfnamefont{D.}~\bibnamefont{Kessler}},
  \bibinfo{author}{\bibfnamefont{J.}~\bibnamefont{Koplik}}, \bibnamefont{and}
  \bibinfo{author}{\bibfnamefont{H.}~\bibnamefont{Levine}},
  \bibinfo{journal}{Physical Review Letters} \textbf{\bibinfo{volume}{51}},
  \bibinfo{pages}{1111} (\bibinfo{year}{1983}).

\bibitem[{\citenamefont{Ben-Jacob et~al.}(1983)\citenamefont{Ben-Jacob,
  Goldenfeld, Langer, and Schon}}]{BENJ83}
\bibinfo{author}{\bibfnamefont{E.}~\bibnamefont{Ben-Jacob}},
  \bibinfo{author}{\bibfnamefont{N.}~\bibnamefont{Goldenfeld}},
  \bibinfo{author}{\bibfnamefont{J.~S.} \bibnamefont{Langer}},
  \bibnamefont{and} \bibinfo{author}{\bibfnamefont{G.}~\bibnamefont{Schon}},
  \bibinfo{journal}{Phys. Rev. Lett.} \textbf{\bibinfo{volume}{51}},
  \bibinfo{pages}{1930} (\bibinfo{year}{1983}).

\bibitem[{\citenamefont{Brower et~al.}(1984)\citenamefont{Brower, Kessler,
  Koplik, and Levine}}]{BROW84}
\bibinfo{author}{\bibfnamefont{R.~C.} \bibnamefont{Brower}},
  \bibinfo{author}{\bibfnamefont{D.~A.} \bibnamefont{Kessler}},
  \bibinfo{author}{\bibfnamefont{J.}~\bibnamefont{Koplik}}, \bibnamefont{and}
  \bibinfo{author}{\bibfnamefont{H.}~\bibnamefont{Levine}},
  \bibinfo{journal}{Phys. Rev. A} \textbf{\bibinfo{volume}{29}},
  \bibinfo{pages}{1335} (\bibinfo{year}{1984}).

\bibitem[{\citenamefont{Campbell and Hanratty}(1982)}]{CAMP82}
\bibinfo{author}{\bibfnamefont{J.~A.} \bibnamefont{Campbell}} \bibnamefont{and}
  \bibinfo{author}{\bibfnamefont{T.~J.} \bibnamefont{Hanratty}},
  \bibinfo{journal}{AlChE Journal} \textbf{\bibinfo{volume}{28}},
  \bibinfo{pages}{988} (\bibinfo{year}{1982}).

\bibitem[{\citenamefont{Campbell and Hanratty}(1983)}]{CAMP83}
\bibinfo{author}{\bibfnamefont{J.~A.} \bibnamefont{Campbell}} \bibnamefont{and}
  \bibinfo{author}{\bibfnamefont{T.~J.} \bibnamefont{Hanratty}},
  \bibinfo{journal}{AlChE Journal} \textbf{\bibinfo{volume}{29}},
  \bibinfo{pages}{215} (\bibinfo{year}{1983}).

\bibitem[{\citenamefont{Ch\'ezy}(1776)}]{CHEZ76}
\bibinfo{author}{\bibfnamefont{A.}~\bibnamefont{Ch\'ezy}}
  (\bibinfo{year}{1776}), \bibinfo{note}{file No. 847, Ms. 1915 in the library
  of Ecole des Ponts et Chauss\'ees. English translation in H. Clemens, On the
  origin of the Ch\'ezy formula, Journal Association of Engineering Societies,
  v. 18, pp. 363-369, (1897).}

\bibitem[{\citenamefont{Saint-Venant}(1871)}]{VENA71}
\bibinfo{author}{\bibfnamefont{B.~D.} \bibnamefont{Saint-Venant}},
  \bibinfo{journal}{Comptes Rendus Acad\'emie des Sciences, Paris, Tome 73,
  July}  (\bibinfo{year}{1871}).

\bibitem[{\citenamefont{Kolmogorov}(1941)}]{KOLM41}
\bibinfo{author}{\bibfnamefont{A.~N.} \bibnamefont{Kolmogorov}},
  \bibinfo{journal}{Dokl. Akad. Nauk. SSSR} \textbf{\bibinfo{volume}{30}},
  \bibinfo{pages}{299} (\bibinfo{year}{1941}), \bibinfo{note}{[English
  translation in Proc. R. Soc. London Ser. A 434 (1991)]}.

\bibitem[{\citenamefont{Sreenivasan}(1999)}]{SREE99}
\bibinfo{author}{\bibfnamefont{K.~R.} \bibnamefont{Sreenivasan}},
  \bibinfo{journal}{Rev. Mod. Phys.} \textbf{\bibinfo{volume}{71}},
  \bibinfo{pages}{S383} (\bibinfo{year}{1999}).

\bibitem[{\citenamefont{Dressler}(1978)}]{DRES78}
\bibinfo{author}{\bibfnamefont{R.~F.} \bibnamefont{Dressler}},
  \bibinfo{journal}{J. Hydraul. Res.} \textbf{\bibinfo{volume}{16}},
  \bibinfo{pages}{205} (\bibinfo{year}{1978}).

\bibitem[{\citenamefont{Sivakumaran et~al.}(1981)\citenamefont{Sivakumaran,
  Hosking, and Tingsanchali}}]{SIVA81}
\bibinfo{author}{\bibfnamefont{N.~S.} \bibnamefont{Sivakumaran}},
  \bibinfo{author}{\bibfnamefont{R.}~\bibnamefont{Hosking}}, \bibnamefont{and}
  \bibinfo{author}{\bibfnamefont{T.}~\bibnamefont{Tingsanchali}},
  \bibinfo{journal}{J. Fluid Mech.} \textbf{\bibinfo{volume}{111}},
  \bibinfo{pages}{411} (\bibinfo{year}{1981}).

\bibitem[{\citenamefont{Friedman}(1970)}]{FRIE70}
\bibinfo{author}{\bibfnamefont{I.}~\bibnamefont{Friedman}},
  \bibinfo{journal}{Geochimica et Cosmochimica Acta}
  \textbf{\bibinfo{volume}{34}}, \bibinfo{pages}{1303} (\bibinfo{year}{1970}).

\bibitem[{\citenamefont{Pentecost}(1990)}]{PENT90}
\bibinfo{author}{\bibfnamefont{A.}~\bibnamefont{Pentecost}},
  \bibinfo{journal}{Geol. Mag.} \textbf{\bibinfo{volume}{127}},
  \bibinfo{pages}{159} (\bibinfo{year}{1990}).

\bibitem[{\citenamefont{Liu and Goldenfeld}(1988)}]{Liu88}
\bibinfo{author}{\bibfnamefont{F.}~\bibnamefont{Liu}} \bibnamefont{and}
  \bibinfo{author}{\bibfnamefont{N.}~\bibnamefont{Goldenfeld}},
  \bibinfo{journal}{Physical Review A} \textbf{\bibinfo{volume}{38}},
  \bibinfo{pages}{407} (\bibinfo{year}{1988}).

\bibitem[{\citenamefont{Barber et~al.}(1987)\citenamefont{Barber, Barbieri, and
  Langer}}]{barber1987dds}
\bibinfo{author}{\bibfnamefont{M.}~\bibnamefont{Barber}},
  \bibinfo{author}{\bibfnamefont{A.}~\bibnamefont{Barbieri}}, \bibnamefont{and}
  \bibinfo{author}{\bibfnamefont{J.}~\bibnamefont{Langer}},
  \bibinfo{journal}{Physical Review A} \textbf{\bibinfo{volume}{36}},
  \bibinfo{pages}{3340} (\bibinfo{year}{1987}).

\bibitem[{\citenamefont{Pieters and Langer}(1986)}]{pieters1986nds}
\bibinfo{author}{\bibfnamefont{R.}~\bibnamefont{Pieters}} \bibnamefont{and}
  \bibinfo{author}{\bibfnamefont{J.}~\bibnamefont{Langer}},
  \bibinfo{journal}{Physical Review Letters} \textbf{\bibinfo{volume}{56}},
  \bibinfo{pages}{1948} (\bibinfo{year}{1986}).

\bibitem[{\citenamefont{Gonz\'alez-Cinca
  et~al.}(2001)\citenamefont{Gonz\'alez-Cinca, Ram\'irez-Piscina, Casademunt,
  and Hern\'andez-Machado}}]{GONZ01}
\bibinfo{author}{\bibfnamefont{R.}~\bibnamefont{Gonz\'alez-Cinca}},
  \bibinfo{author}{\bibfnamefont{L.}~\bibnamefont{Ram\'irez-Piscina}},
  \bibinfo{author}{\bibfnamefont{J.}~\bibnamefont{Casademunt}},
  \bibnamefont{and}
  \bibinfo{author}{\bibfnamefont{A.}~\bibnamefont{Hern\'andez-Machado}},
  \bibinfo{journal}{Phys. Rev. E} \textbf{\bibinfo{volume}{63}},
  \bibinfo{pages}{051602} (\bibinfo{year}{2001}).

\bibitem[{\citenamefont{Ben-Jacob et~al.}(1984)\citenamefont{Ben-Jacob,
  Goldenfeld, Kotliar, and Langer}}]{benjacob1984psd}
\bibinfo{author}{\bibfnamefont{E.}~\bibnamefont{Ben-Jacob}},
  \bibinfo{author}{\bibfnamefont{N.}~\bibnamefont{Goldenfeld}},
  \bibinfo{author}{\bibfnamefont{B.}~\bibnamefont{Kotliar}}, \bibnamefont{and}
  \bibinfo{author}{\bibfnamefont{J.}~\bibnamefont{Langer}},
  \bibinfo{journal}{Physical Review Letters} \textbf{\bibinfo{volume}{53}},
  \bibinfo{pages}{2110} (\bibinfo{year}{1984}).

\bibitem[{\citenamefont{Kessler et~al.}(1985)\citenamefont{Kessler, Koplik, and
  Levine}}]{kessler1985gmi}
\bibinfo{author}{\bibfnamefont{D.}~\bibnamefont{Kessler}},
  \bibinfo{author}{\bibfnamefont{J.}~\bibnamefont{Koplik}}, \bibnamefont{and}
  \bibinfo{author}{\bibfnamefont{H.}~\bibnamefont{Levine}},
  \bibinfo{journal}{Physical Review A} \textbf{\bibinfo{volume}{31}},
  \bibinfo{pages}{1712} (\bibinfo{year}{1985}).

\bibitem[{\citenamefont{Bales and Zangwill}(1990)}]{bales1990mit}
\bibinfo{author}{\bibfnamefont{G.}~\bibnamefont{Bales}} \bibnamefont{and}
  \bibinfo{author}{\bibfnamefont{A.}~\bibnamefont{Zangwill}},
  \bibinfo{journal}{Physical Review B} \textbf{\bibinfo{volume}{41}},
  \bibinfo{pages}{5500} (\bibinfo{year}{1990}).

\end{thebibliography}

\end{document}